\documentclass[twocolumn]{aastex7}
\usepackage{amsmath}

\begin{document}

\title{Fast X-ray Transients produced by Off-axis Jet-Cocoons from Long Gamma-Ray Bursts}

\author[orcid=0000-0001-5751-633X]{Jian-He Zheng} 
\affiliation{School of Astronomy and Space Science, Nanjing University, Nanjing 210023, People’s Republic of China}
\affiliation{Department of Astronomy and Theoretical Astrophysics Center, University of California, Berkeley, CA 94720, USA}
\email{mg21260020@smail.nju.edu.cn}  

\author[0000-0002-1568-7461]{Wenbin Lu}
\affiliation{Department of Astronomy and Theoretical Astrophysics Center, University of California, Berkeley, CA 94720, USA}
\email{wenbinlu@berkeley.com}

\begin{abstract}

Fast X-ray transients (FXTs) have been detected for over a decade, yet their origins are still enigmatic. The observed association between FXTs and broad-lined Type Ic supernovae (SNe Ic-BL) suggests that some may share the same progenitor with Long Gamma-Ray Bursts. 
In this work, we numerically simulate the long-term evolution of a relativistic jet propagating from inside the progenitor star up to the photon diffusion radius of the cocoon. Then we post-process the hydrodynamic results and calculate the cocoon cooling emission for various viewing angles from the jet axis. 
We find that, for viewing angles $\theta_{\rm v}=10^{\circ}$-$20^{\circ}$, the off-axis cocoon emission can produce FXTs with luminosity $L_{\rm X}\simeq 10^{47-48} {\rm\, erg\,s^{-1}}$ and duration $t_{\rm X}\simeq 10$-$100\,$s. The observed spectra are quasi-thermal with the peak energy $E_{\rm peak}\simeq0.8$ keV. 
These properties naturally explain observational features of { a fraction of FXTs}, including their high luminosity, soft spectra, and lack of gamma-ray counterparts. The Rayleigh-Jeans tail of the FXT spectra extends to the UV, producing an early UV flash simultaneously. As the cocoon expands and cools, the emission peak shifts to UV and optical bands, resulting in a bright optical plateau lasting for $\sim1$ day with color temperature $T_{\rm  UV/opt} \simeq (1{-}3)\times10^{4}\,$K { and bolometric luminosity $L_{\rm bol}\simeq10^{41-42} {\rm\, erg\,s^{-1}}$}, before the emergence of supernova emission. Although our model underpredicts the UV/optical luminosity at $\sim1$ day { for some events (e.g. EP 240414a)}, it still provides useful diagnostics for identifying the origins of FXTs.


\end{abstract}


\section{Introduction} 

Fast X-ray transients (FXTs) are soft X-ray bursts with a wide range of timescales $10^{1}-10^{4}$s. Some of them are incidentally discovered by Swift observations and Chandra surveys \citep[e.g.,][]{Soderberg2008Natur.453..469S,Jonker2013ApJ...779...14J,Glennie2015MNRAS.450.3765G,Luo2017ApJS..228....2L,Bauer2017MNRAS.467.4841B,Xue2019Natur.568..198X}. Then, the systematic search in the archival data uncovered tens of FXTs in Chandra, XMM-Newton and BeppoSAX datasets\citep[e.g.,][]{Yang2019MNRAS.487.4721Y,Alp2020ApJ...896...39A,Quirola2022A&A...663A.168Q,Quirola2023A&A...675A..44Q,Zand2025arXiv251216845I}.
Numerous models have been proposed to explain their origins, including stellar flares \citep{Glennie2015MNRAS.450.3765G}, supernova shock breakouts \citep{Soderberg2008Natur.453..469S}, Long Gamma-ray Burst (LGRB) jets \citep{Quirola2022A&A...663A.168Q}, and nascent millisecond magnetars \citep{Xue2019Natur.568..198X}. However, the nature of FXTs in these archival discoveries remains elusive because, for most of them, multi-wavelength follow-ups were unavailable to provide the identifications.

This field has been revolutionized since the launch of the Einstein Probe (EP) satellite. 
Two instruments are onboard the EP. The Wide-field X-ray Telescope (WXT) has a large field of view, high sensitivity, and arcminute-level localization capability, enabling high cadence sky scans and rapid localizations of FXTs \citep{Yuan2022hxga.book...86Y}. The Follow-up X-ray Telescope provides rapid follow-ups and arcsecond-level localizations. These capabilities allow us to discover many FXTs and organize rapid follow-ups in optical and radio bands. 

Numerous FXTs have been discovered since the operation of EP in 2024, most of which appear as ``orphan'' X-ray transients without gamma-ray counterparts \citep[e.g.,][]{Zhang2025ApJ...987L..38Z}. Despite the fact that 95\% of the EP samples have gamma-ray monitor coverage during the X-ray triggers \citep{Zhang2025ApJ...987L..38Z}, 78\% of FXTs have no gamma-ray counterparts. A subset of the FXTs is associated with LGRBs, implying some FXTs originate from relativistic jets launched in core-collapse supernovae \citep[e.g., GRB 240219A, GRB 240315A,][]{Yin2024ApJ...975L..27Y,Levan2025NatAs...9.1375L,Liu2025NatAs...9..564L}. The fact that most FXTs have no or very faint gamma-ray signals disfavors a standard on-axis GRB jet origin for the majority of them.

However, several FXTs are confirmed to have broad-lined Type Ic supernova (SN Ic-BL) counterparts in optical follow-up observations \citep[EP 240414a, EP 250108a,][]{vanDalen2025ApJ...982L..47V,Srivastav2025ApJ...978L..21S,Sun2025NatAs...9.1073S,Rastinejad2025ApJ...988L..13R, Li2025arXiv250417034L}, and some show SN Ic-BL candidates \citep[EP 241021a][]{Quirola2025MNRAS.tmp.1943Q}. The association between SNe Ic-BL and LGRBs has been confirmed in the past two decades \citep{Woosley2006ARA&A..44..507W,Hjorth2012grb..book..169H}. Furthermore, the radio follow-up within 10-100 days reveals bright and long-lived radio afterglows for these FXTs, which are comparably bright with the afterglow from ordinary LGRBs on similar timescales \citep[EP240414a, EP241021a,][]{Bright2025ApJ...981...48B,Yadav2025ApJ...995..216Y}. 

{
The presence of bright radio afterglows implies that these FXTs are powered by mildly relativistic outflows, suggesting a connection to jet-driven explosions. Within this framework, the X-ray prompt emission may originate from the dissipation of a weak jet \citep{Sun2025NatAs...9.1073S,Hamidani2025ApJ...986L...4H} or cocoon cooling, while the subsequent rebrightening of some events at a few days can be attributed to shock cooling \citep{Sun2025NatAs...9.1073S}, energy injected afterglows \citep[refreshed shocks,][]{Srivastav2025ApJ...978L..21S,Busmann2025A&A...701A.225B}, or off-axis afterglows \citep{Zheng2025ApJ...985...21Z}.
}

Motivated by these observations, an off-axis jet-cocoon model was proposed to explain the complete evolution of the SN Ic-BL-associated FXTs \citep{Zheng2025ApJ...985...21Z}.
Cocoon formation is inevitable since the relativistic jet strongly interacts with the stellar envelope during its propagation within the star, which has been studied theoretically \citep[e.g.,][]{Ramirez2002MNRAS.337.1349R,Bromberg2011ApJ...740..100B,Nakar2017ApJ...834...28N,DeColle2018MNRAS.478.4553D,Hamidani2023MNRAS.524.4841H} and observationally \citep{Izzo2019Natur.565..324I}. Roughly speaking, the cocoon has two components: a mildly relativistic inner cocoon (shocked jet) and a non-relativistic outer cocoon (shocked stellar material). After the jet successfully breaks out of the star, the cocoon expands and eventually releases its thermal energy through cooling emission. The outer cocoon produces bright emission in the ultraviolet and optical bands, whereas the emission from the inner cocoon is relativistically boosted to soft X-ray bands \citep{Nakar2017ApJ...834...28N, DeColle2022MNRAS.512.3627D}.

Analytically, the simplified two-component cocoon model can qualitatively explain the behaviors of the SN Ic-BL-associated FXTs, such as EP240414a \citep{Zheng2025ApJ...985...21Z,Hamidani2025ApJ...986L...4H}. However, such analytical studies do not include the realistic angular profile for the cocoon, which has been obtained from hydrodynamic jet-cocoon simulations \citep[e.g.,][]{Zhang2003,Gottlieb2020MNRAS.498.3320G} but were not directly used to compute the X-ray lightcurves and spectra for different viewing angles. The only exception is the work by \citet{DeColle2018MNRAS.478.4553D} who used a post-processing technique to compute the X-ray lightcurves and spectra from the cocoon emission, but they only considered an on-axis viewing angle at which the cocoon emission will likely be overwhelmed by the prompt emission and the afterglow (and hence this does not apply for FXTs without gamma-ray counterparts).

To overcome these limitations, we carry out both hydrodynamical simulations and post-processing radiative transfer calculations in this work, presenting self-consistent cocoon cooling lightcurves for arbitrary viewing angles. In this new approach, the cocoon is no longer divided into two (i.e., the inner and outer) components but treated as a jet-cocoon system with a continuous angular profile for the energy and Lorentz factor obtained from simulations. We show that this framework naturally reproduces the X-ray lightcurves observed in many FXTs, supporting that some FXTs originate from LGRB progenitors.

A remaining puzzle is that our model significantly under-predicts the luminosity of the thermal UV/optical emission on a timescale of $1\rm\, d$. Despite the discrepancy, we argue that the most likely origin of the early-time, thermal UV/optical emission is the outer regions of the cocoon. Future work with additional jet components and/or an extended stellar envelope may resolve this discrepancy. We also argue that this discrepancy is unlikely to affect the robustness of our predictions for the X-ray emission.

The paper is organized as follows: In Section \ref{sec:methods}, we describe our simulation setups and post-processing methods. Then, we show lightcurves of cooling emissions from the cocoon in Section \ref{sec:res}, including the X-ray, UV, and optical emissions. The discussion and summary are in Sections \ref{sec:dis} and \ref{sec:sum}. Throughout this paper, we use CGS units and ignore cosmological redshift factors except for a few cases (i.e., the observable quantities are expressed in the host galaxy's rest frame).

\section{Methods and Initial Setups}
\label{sec:methods}
\subsection{Hydrodynamics setups}

We aim at capturing the evolution of the jet cocoon system from the jet launching region up to the photon diffusion radius of $\mathcal{O}(10^{14})\rm\, cm$ \citep{Zheng2025ApJ...985...21Z}, which corresponds to an expansion time of $\mathcal{O}(10^4)\rm\, s$ in the lab frame. As full 3D simulations of such long-term jet-cocoon evolution are computationally expensive, we instead perform the 2D axisymmetric simulations using the public code \texttt{PLUTO} v4.4 \citep{Mignone2007ApJS..170..228M}. The 2D setup allows us to explore the parameter space with different jet properties and progenitor star radii. The parameters of all models are listed in Table \ref{tab:param}.

For progenitors of LGRBs, we consider the Wolf-Rayet stellar models developed by \cite{Woosley2006Progenitor}. The pre-supernova mass of the star is assumed to be $M_{\star}=10M_{\odot}$, corresponding to the 12TH model in their paper. The density profile of the star outside the $10^{9}\,$cm (inner boundary of our simulation) can be described by the analytical function $\rho(r)=\rho_0r^{-2}(1-r/R_{\star})^3$, where $R_{\star}$ is the stellar radius and $\rho_0$ is the normalization factor. { We adopt this analytical function in our simulations.}

Outside the stellar surface, the ambient medium is assumed to be shaped by the progenitor's stellar wind, characterized by a mass loss rate of $\dot{M}=10^{-5}M_{\odot}{\rm yr^{-1}}$ and a velocity of $v_{\rm w}=1000\, {\rm km\,s^{-1}}$. Due to its low density, the stellar wind does not affect our results.

Our simulation results are also insensitive to the pressure profiles of the star and the ambient wind, because the total energy is dominated by the rest-mass energy.
For this reason, we set the pressure of the stellar material and the ambient wind as $p = 10^{-5}\rho c^2$. By changing the pressure normalization by a factor of 10 and seeing negligible differences, we have verified that this choice of pressure profile does not affect our results.

All simulations are carried out in spherical coordinates ($r,\theta$) using the relativistic hydrodynamics (RHD) module of \texttt{PLUTO}. We employ the HLL Riemann solver together with third-order Runge–Kutta (RK3) time stepping integration, using the parabolic reconstruction method and harmonic mean van Leer limiter. The equation of state is that of ideal gas with an adiabatic index of $4/3$, appropriate for the radiation-dominated jet-cocoon system. Our lightcurve calculations only rely on the data in regions where the photon diffusion time is (at least marginally) longer than the dynamical expansion time, so an adiabatic equation of state is justified.

The radial grid is divided into three logarithmically uniform patches. The first patch extends from the inner boundary $r_{\rm in}=10^{9}\,$cm to $10^{11}\,$cm with 400 cells. The second patch covers the range from $10^{11}\,$cm to $10^{12}\,$cm with 400 cells. The last patch stretches from $10^{12}\,$cm to the outer boundary $r_{\rm out}= 7\times10^{14}\,$cm with 1200 cells. 
The angular grids consist of two uniform patches, which are 200 cells from 0 to $30^{\circ}$ and another 200  cells from $30^{\circ}$ to $90^{\circ}$. The full grid therefore contains $2000\times400$ cells in $r\times \theta$. In Appendix \ref{app:conver}, we verify that this resolution reaches convergence in our setups. 

\begin{table}[]
    \centering
    \begin{tabular}{c|c|c|c}
    \hline
        Models & $L_{\rm j}[{\rm erg\,s^{-1}}]$  & $\theta_{\rm j,0}[^{\circ}]$ & $R_{\star}[10^{11}{\rm cm}]$      \\
    \hline
        Canonical ($Lc$) &      $1.5\times10^{50}$   &  5   &   1  \\
        Wide angle ($Lw$) &     $1.5\times10^{50}$   &  10  &   1  \\
        Inflated star ($LI$) &  $1.5\times10^{50}$   &  5   &   4  \\
        Low energy ($Llow$) &   $1.5\times10^{49}$   &  5   &   1   \\
    \hline
    \end{tabular}
    \caption{Simulation model parameters: $L_{\rm j}$ is the two-sided jet luminosity, $\theta_{\rm j,0}$ is the initial half opening angle of the jet, $R_{\star}$ is the radius of the progenitor, and the jet duration is fixed to be $t_{\rm j}=20\,$s for all models. 
    }
    \label{tab:param}
\end{table}

We inject the conical jet from the nozzle located at $10^{9}\,$cm with its velocity directed radially outward. Since the radius of the GRB progenitor is $R_{\star}\sim10^{11}\,$cm, the injected radius corresponds to $0.01R_{\star}$, meaning that subsequent evolution of the jet-cocoon system is not affected by the nozzle location \citep{Harrison2018MNRAS.477.2128H,Gottlieb2020MNRAS.498.3320G}.
The angular profiles of the jet energy and velocity are described by ${\rm cosh^{-8}(\theta/\theta_{\rm j,0})}$, which is similar to the top-hat shape. The jet structure is thus roughly uniform within the half opening angle $\theta_{\rm j,0}$ and declines steeply beyond $\theta_{\rm j,0}$. The jet is initially hot and is launched with an initial Lorentz factor $\Gamma_0$. We adopt $\Gamma_0 = 0.7 \theta_{\rm j,0}^{-1}$, motivated by the fact that only material within an angle of $\sim\!\Gamma^{-1}$ is causally connected, and the factor of 0.7 is obtained from previous studies \citep{Mizuta2013ApJ...777..162M}. We adopt the outflow boundary condition at the outer radial boundary at $7\times10^{14}\rm\,cm$, and reflective boundaries at both the pole and the equator of the angular grid.

The maximum terminal 4-velocity of the jet on the polar axis is set to be $u_{\infty}\equiv\sqrt{\Gamma^2_0h'^2_0-1}=20$, where $h'_0=1+4p_{\rm j,0}'/(\rho'_{\rm j,0}c^2)$ is the dimensionless specific enthalpy. We use a modest terminal 4-velocity that is much smaller than that of the canonical GRB jet core because we focus on the off-axis ($\gtrsim10^{\circ}$) emission from cocoon cooling instead of the prompt gamma-ray emission from the jet core. A smaller terminal velocity relaxes the requirement for resolution and computational expenses.

Physically, the terminal velocity of the jet-cocoon system decreases rapidly from the jet core, resulting in $u_{\infty}\lesssim10$ for $\theta\gtrsim10^{\circ}$ \citep{Gottlieb2020MNRAS.498.3320G}. This means that our chosen core $u_\infty=20$ does not bias the cocoon emission at large viewing angles ($\gtrsim10^\circ$) because the relevant emitting material has $u_\infty\lesssim 10$ anyway. In Appendix \ref{app:conver}, we test two other choices of $u_\infty=15$ and $30$, which demonstrate that our choice of $u_\infty=20$ does not substantially affect the predicted lightcurves.

We assume that the power injection remains constant for a duration of $t_{\rm j}\approx20\,$s in the lab frame and that the injection cuts off after $t_{\rm j}$. At $t>t_{\rm j}$, we adopt an exponential cutoff for the jet velocity and pressure, which are given by
\begin{equation}
(\beta, p') = 
\left\{
\begin{array}{ll}
(\beta_0,\, p'_{\rm j,0}), & t \le t_{\rm j}, \\[6pt]
(\beta_0,\, p'_{\rm j,0})\, (e^{\,1 - t/t_{\rm j}}+10^{-7}) , & t > t_{\rm j} ,
\end{array}
\right.
\end{equation}
where $p'_{\rm j,0}$ is the initial pressure in the comoving frame and $\beta_0=\sqrt{1-\Gamma^{-2}_0}$ is the initial velocity in the lab frame. The density boundary condition near the nozzle is kept constant during and after the engine duration. Because of low velocity and pressure, material injected after $t_{\rm j}$ is non-relativistic and has no impact on the jet-cocoon system that has already broken out from the stellar surface. The factor $10^{-7}$ is adopted to avoid numerical errors caused by the extremely small pressure values. The exponential cutoff not only speeds up the simulation (by improving time stepping by a factor of $\geq10$) but also improves numerical stability.


We run the simulations up to a maximum lab-frame time of $t_{\rm max} = 2\times10^4\rm\,s$. By this stage, the jet-cocoon system has already entered the homologous expansion phase. 
Afterwards, we extrapolate the evolution by assuming homologous expansion, such that each fluid element expands as $r\propto t$, while the existing density and pressure profiles are preserved in shape and evolve as $\rho \propto r^{-3}$ and $p\propto \rho^{4/3}\propto r^{-4}$.

\subsection{Post-processing}
After the hydrodynamic simulation is completed, we apply the following post-processing procedure to compute the lightcurves of the cocoon cooling emission. Our method is qualitatively similar to that adopted by \citet{DeColle2018MNRAS.478.4553D}, with the quantitative differences : (1) we explicitly take into account radiative diffusion between the photon diffusion radius $r_{\rm diff}$ and the photosphere $r_{\rm ph}$ (see below) and (2) we consider an arbitrary observer's viewing angle $\theta_{\rm v}$. The difference between $r_{\rm diff}$ and $r_{\rm ph}$ is negligible for the mildly relativistic X-ray emitting gas, but the difference is important for the sub-relativistic gas that dominates the late-time (hours--days) emission in the UV/optical bands.

\begin{figure*}
    \centering
    \includegraphics[width=\textwidth]{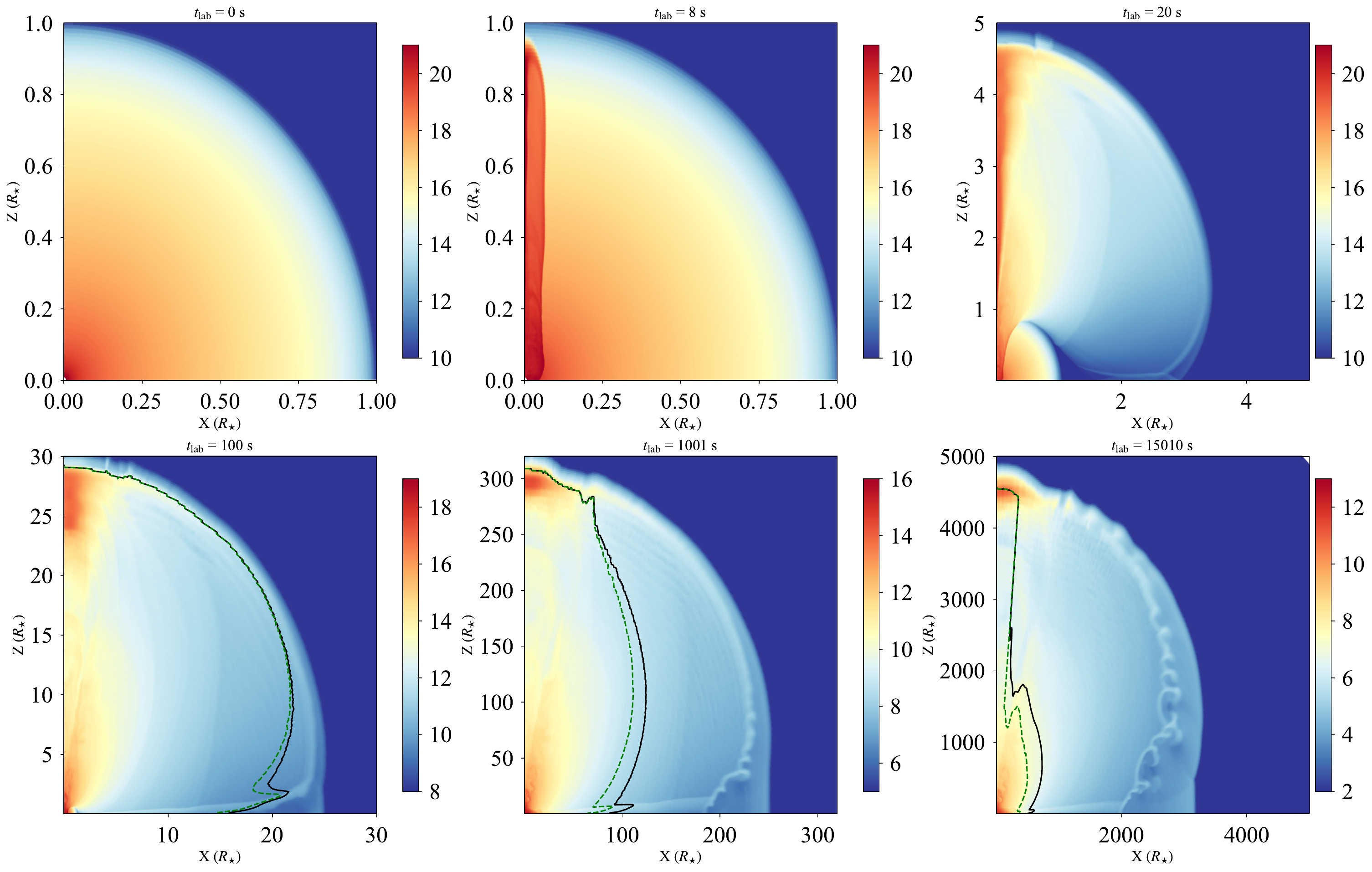}
    \caption{Energy density maps, the sum of kinetic energy density $\Gamma(\Gamma-1)\rho'c^2$ and thermal energy density $\Gamma^2 e'$ of the canonical model (Lc). 
    Different panels correspond to different lab frame times.
    The colorbar corresponds to the logarithmic energy density in unit of ${\rm erg\,{cm}^{-3}}$. The black solid lines and green dashed lines in the bottom panel are photospheric radius $r_{\rm ph}$ and diffusion radius $r_{\rm diff}$, respectively.
    }
    \label{fig:Engden}
\end{figure*}

At a given lab frame time $t_{\rm lab}$, to calculate the radial position of the diffusive shells $r_{\rm diff}(\theta)$ along a given polar angle $\theta$, we first calculate the optical depth of the jet-cocoon system \textit{along the radial direction}, which is 
\begin{equation}
    \tau(r,\theta)=\int^{\infty}_{r}\kappa_{\rm R}\Gamma(r,\theta)\rho'(r,\theta) dr,
\end{equation}
where $\kappa_{\rm R}\approx0.2\,{\rm cm^{2}g^{-1}}$ is the Rosseland-mean opacity (approximated by electron scattering for hydrogen-poor gas composition) and $\rho'$ is the density in the comoving frame (such that the lab-frame density is $\Gamma \rho'$). Based on the optical depth, we define the photospheric radius along a given polar angle $r_{\rm ph}(\theta)$ by
\begin{equation}
    \tau(r_{\rm ph},\theta)=1 \ \ \Rightarrow \ \ r_{\rm ph}(\theta).
\end{equation}
{ We note that this approximation neglects the limb-darkening effect.}
Photons begin to diffuse out when the optical depth of fluid elements drops below $\tau\approx\beta^{-1}$. We therefore solve for the diffusion radius $r_{\rm diff}(\theta)$ along each polar angle $\theta$ according to
\begin{equation}
    \tau(r_{\rm diff},\theta)\approx [\beta(r_{\rm diff},\theta)]^{-1} \ \ \Rightarrow  \ \ r_{\rm diff}(\theta).
\end{equation}
For mildly relativistic emitting gas at small polar angles, we expect $r_{\rm diff}\approx r_{\rm ph}$, whereas for sub-relativistic gas at larger polar angles, the difference between these two radii is not negligible (see Figure \ref{fig:Engden}). { The definitions of $r_{\rm ph}$ and $r_{\rm diff}$ remain valid as long as the cocoon interior is optically thick. In our simulations, the cocoon interior is optically thick up to $t_{\rm lab}\sim 3$ days.
}

We assume that the energy flux from radiative diffusion is dominated by the component along the radial direction, which can be estimated by
\begin{equation}
    F'_{\rm diff}(\theta) =\frac{c}{3\rho' \kappa_{\rm R}}\frac{\partial e'_{\rm rad}}{\partial r'}  \approx \frac{ce'_{\rm rad}(r_{\rm diff},\theta)}{\tau(r_{\rm diff},\theta)},
\end{equation}
where $e'_{\rm rad}(r_{\rm diff},\theta)\approx 3p'(r_{\rm diff},\theta)$ is the radiative energy density at the diffusion radius (for radiation-dominated gas). In the above estimate, we have taken an approximation $\tau \approx 3\rho' \kappa_{\rm R}  e'_{\rm rad}/|\partial e'_{\rm rad}/\partial r'|$, which holds as the energy density has a roughly power-law radial profile. To reduce numerical fluctuations, we adopt the average of three neighboring grid points along the radial direction when computing the flux $F'_{\rm diff}(\theta)$.

When the photon diffusion timescale is shorter than the dynamical expansion timescale (or $\tau<\beta^{-1}$), we can ignore adiabatic losses at $r > r_{\rm diff}$, so the diffusive luminosity is approximately conserved during the photon propagation from the diffusion radius $r_{\rm diff}(\theta)$ to the photosphere $r_{\rm ph}(\theta)$, leading to the flux at the photosphere along a given polar angle $\theta$,
\begin{equation}
    F'_{\rm ph}(\theta)\approx F'_{\rm diff}(\theta)\left[ \frac{r_{\rm ph}(\theta)}{r_{\rm diff}(\theta)}\right]^{-2}.
\end{equation}
Below the photosphere, we assume that the radiation field has a blackbody spectral shape and is isotropic in the comoving frame of a given fluid element. We also assume that the temperature of the escaping radiation near the photosphere $T_{\rm ph}'(\theta)$ is determined near the diffusion radius, i.e.,
\begin{equation}
    T_{\rm ph}'(\theta)\approx T'_{\rm diff}(\theta) =\left[\frac{e'_{\rm rad}(r_{\rm diff},\theta)}{a}\right]^{1/4} ,
\end{equation}
where $a$ is the radiation density constant. This is expected as the cocoon's opacity is strongly dominated by scattering (as opposed to bound-free and free-free absorption), resulting in the thermalized radius close to the diffusion radius $r_{\rm diff}$ (smaller than the photospheric radius $r_{\rm ph}$). The blackbody spectral shape is only valid if the gas and radiation in a given fluid element reach local thermodynamic equilibrium (LTE) \textit{when the cocoon was initially created} --- such a blackbody spectrum will be preserved during the subsequent adiabatic expansion and radiative diffusion as long as the opacity is nearly gray. In Appendix \ref{app:LTE}, we verify that nearly all fluid elements inside the cocoon satisfy the LTE condition, i.e., sufficient photons are produced by free-free emission within a dynamical timescale (or effective optical depth greater than unity) before the breakout.


The specific intensity in the comoving frame of the fluid elements near the photosphere is
\begin{equation}
    \label{eq:BB}
    I'_{\nu}(\theta)\approx \frac{F'_{\rm ph}(\theta)}{\sigma_{\rm SB} T'^4_{\rm diff}(\theta)}B_{\nu}[T'_{\rm diff}(\theta)],
\end{equation}
where $\sigma_{\rm SB}$ is Stefan-Boltzmann constant, and $B_{\nu}(T)$ is the Planck function. We assume that the specific intensity is isotropic over an outward-facing hemisphere, so the total energy flux in the radial direction is given by $\int_0^\infty d\nu' \int_{\theta'=0}^{\pi/2} I'_{\nu'} \mu' d\Omega' = \pi \int I'_{\nu'}d \nu' = F'_{\rm ph}(\theta)$, where $\mu'$ is the usual projection factor onto the radial direction.

Once we have computed the specific intensity near the photosphere, the observed flux is then obtained by an integral over the relativistic Equal Arrival Time Surface (EATS). Using Lorentz transformation with the Doppler factor $\mathcal{D}(\theta_{\rm v},\theta,\varphi)=[\Gamma(1-\beta\cos\chi)]^{-1}$,
\begin{equation}
    I_{\nu}=\mathcal{D}^3 I'_{\nu'},
\end{equation}
We obtain the flux density in the observer's frame by integrating over the entire photosphere
\begin{equation}\label{eq:observed_flux}
    F_{\rm\nu, obs}=\frac{1}{d^2}\int^{\pi/2}_{0}\int^{2\pi}_0\mathcal{D}^3(\theta_{\rm v},\theta,\varphi)I'_{\nu'}(\theta)|\cos{\chi'}|r^2_{\rm ph}(\theta)d\Omega,
\end{equation}
where $d$ is the (luminosity) distance, $\nu =\mathcal{D}\nu^{\prime}$ is the frequency in the observer's frame, and the integral is over {the solid angle in the northern hemisphere} ($0<\theta<\pi/2$). { We show that, due to symmetry of the system, the integral over only one hemisphere (on the side of the observer) is a reasonable approximation for our simulations in Appendix \ref{app:sphere}. } In the above equations, $\chi$ is the angle between the radial direction and the observer's line of sight (LOS),
\begin{equation}
    \label{eq:project}
    \cos\chi =\cos{\theta_{\rm v} }\cos\theta  + \sin{\theta_{\rm v}}\sin\theta\cos\varphi,
\end{equation}
where $\theta_{\rm v}$ is the observer's viewing angle with respect to the jet axis. 
The factor of $|\cos{\chi'}|$ projects the emitting surface area on the EATS in the direction perpendicular to the LOS in the comoving frame, and $\cos\chi'=(\cos\chi-\beta)/(1-\beta\cos\chi)$ given by the Lorentz transformation. Nominally, the angle appearing in the Doppler factor should be the angle between the gas velocity and LOS, but since gas velocity is dominated by the radial component when radiative diffusion becomes important, we adopt $\chi$ in the Doppler factor for simplicity (with negligible loss in accuracy).

Throughout this paper, we present the observable emission in the form of \textit{isotropic equivalent} spectral luminosity, as given by
\begin{equation}
    L_{\nu} = 4\pi d^2 F_{\rm \nu, obs},
\end{equation}
where $d$ is the distance to the source.

The integral in Eq. (\ref{eq:observed_flux}) implicitly includes different lab frame time $t_{\rm lab}(t_{\rm obs}, \theta)$ for a given observer's time and for contribution near a given polar angle $\theta$. Finally, to convert the observer's time $t_{\rm obs}$ to the lab-frame time $t_{\rm lab}$ used in the simulation, we use the retarded time of a moving emitter
\begin{equation}\label{eq:obstime}
\begin{aligned}
    t_{\rm obs}
    &=t_{\rm lab}- t_{\rm bo}-\frac{(r_{\rm ph}-R_{\star})\cos\chi}{c}   \\
    &\approx(t_{\rm lab}-t_{\rm bo})(1-\beta_{\rm ph}\cos\chi).
\end{aligned}
\end{equation}
Here we define the zero point of the observer's time $t_{\rm obs}=0$ corresponding to the breakout time $t_{\rm bo}$ on the jet axis (measured in the lab frame) when $r_{\rm ph}=R_{\star}$. Due to finite spatial resolution, we do not accurately resolve the $\tau\sim 1$ region with more than a few cells at early time right after the breakout (this is less of an issue when the cocoon has expanded to $r\sim 10^{14}\rm\, cm$ where most emission is produced). The inaccuracy in the photospheric radius can sometimes cause the observer's time $t_{\rm obs}$ to become negative (unphysical). To avoid this issue, we adopt an approximation in Eq. (\ref{eq:obstime}) by assuming that the cocoon photosphere has been expanding at a constant velocity along the radial direction that is equal to the current velocity at the photosphere $\beta_{\rm ph} c$. When the cocoon has expanded to $r\sim 10^{14}\rm\,cm$, the effects of acceleration and lateral expansion are negligible, and our assumption is asymptotically correct $\beta_{\rm ph} c\approx (r_{\rm ph}-R_\star)/(t_{\rm lab}-t_{\rm bo})$. Our approach produces reliable lightcurves at late time (after the first few seconds) and avoids any negative $t_{\rm obs}$ at early time.



\section{Results}
\label{sec:res}

\subsection{Physical Properties of the cocoon}
\label{sec:phy}
The evolution of the jet-cocoon system in our canonical model $Lc$ is shown in Figure \ref{fig:Engden}. The jet head propagates at a sub-relativistic velocity within the star and breaks out from the star at $t_{\rm bo}\approx8\,$s in the lab frame. The cocoon then expands laterally and gradually becomes quasi-spherical. After the jet shuts off at $t_{\rm j}=20\,$s, the additional energy injected from the nozzle is negligible. The jet continues to move outward and eventually forms a thin shell ahead of the jet-cocoon system, while the quasi-spherical cocoon expands radially. 

Material near the jet head spreads laterally and induces some instabilities on the cocoon front. These instabilities are visible as wavy structures in the bottom-right panel of Figure \ref{fig:Engden}. They introduce some small fluctuations on the early optical/UV lightcurves shown in \S\ref{sec:opt}. Once photons within these shells have diffused out, the structures no longer affect the lightcurves.

Cocoon material near the jet axis is mildly relativistic ($\Gamma\sim$ a few) and referred to as the inner cocoon. Due to the relativistic motion and hot temperatures, emission from the inner cocoon is boosted to soft X-ray bands, resulting in bright X-ray emission at early times. 
Later on, the photons carried by the trans-relativistic and non-relativistic material (``outer cocoon'') begin to diffuse out. Therefore, the diffusion time is much longer, and the radiation spectrum has a lower temperature, so we expect that continuous cooling emission in the UV follows the FXT. 

\begin{figure}
    \centering
    \includegraphics[width=0.9\linewidth]{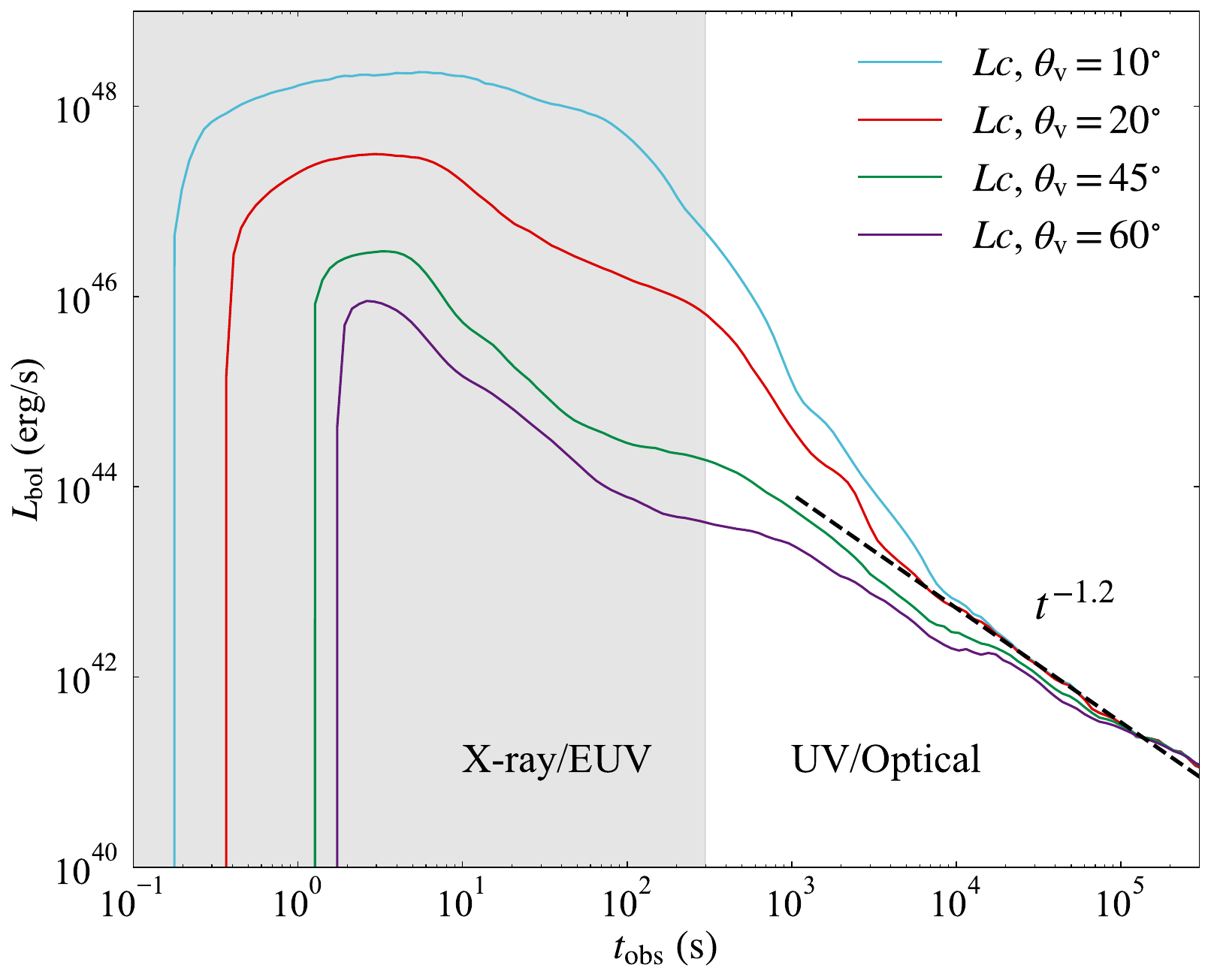}
    \includegraphics[width=0.9\linewidth]{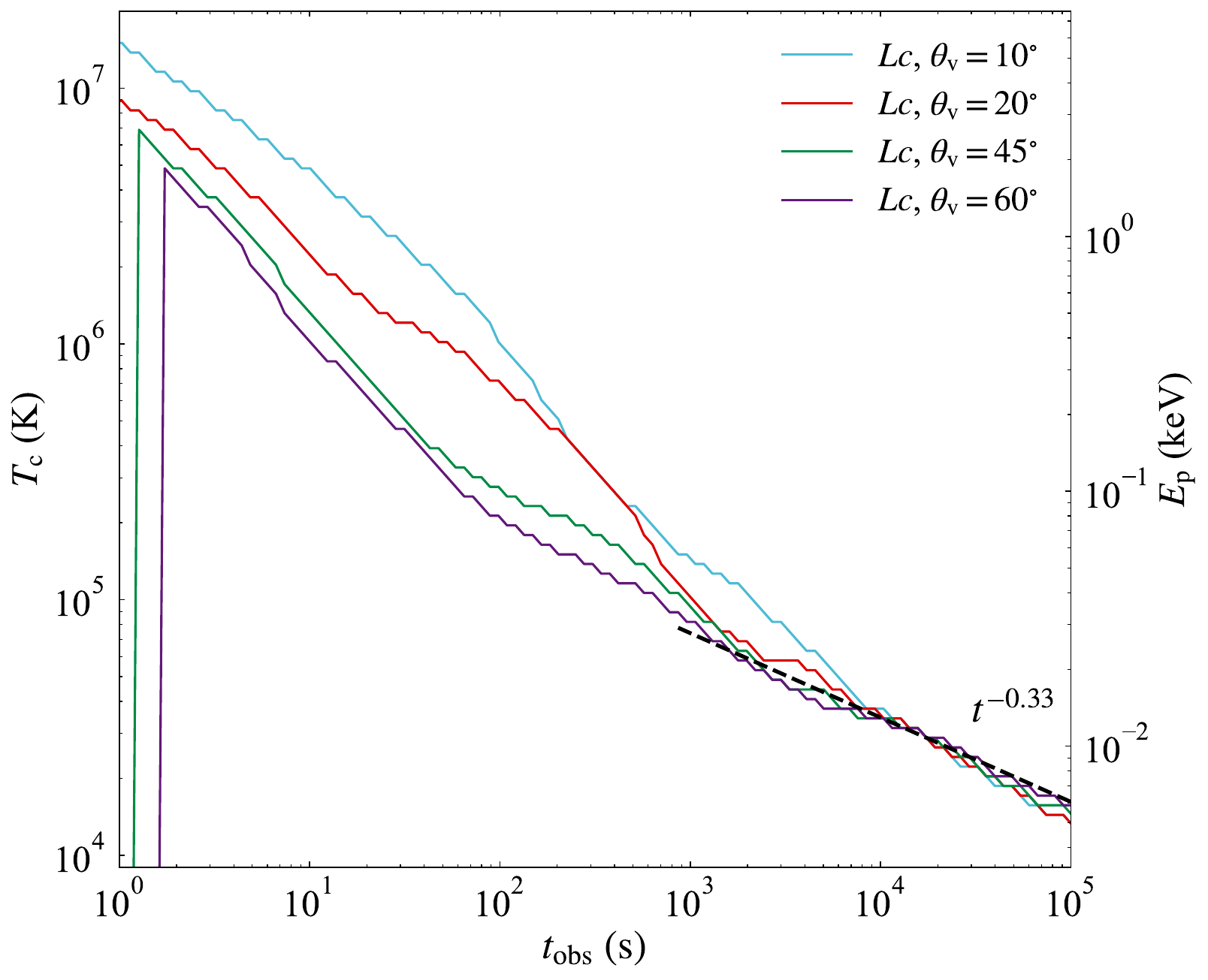}
    \caption{\textit{Upper panel}: Bolometric lightcurves (including all bands, X-ray/UV/optical) for the $Lc$ model. The blue, red, green, and purple lines correspond to the viewing angles of $10^{\circ}$, $20^{\circ}$, $45^{\circ}$, and $60^{\circ}$. Four lightcurves converge at $\sim1$ day. \textit{Lower panel}: The evolution of color temperature for the \textit{Lc} model. The corresponding peak energy $E_{\rm peak}$ is shown on the right axis.
    }
    \label{fig:Lcbol}
\end{figure}

{
As a result of the weaker Doppler boost, the cooling emission eventually becomes quasi-isotropic, and the viewing-angle dependence weakens. This behavior is shown in Figure \ref{fig:Lcbol}, where the bolometric luminosities $L_{\rm bol}=\int L_{\nu}d\nu$ (including optical/UV/X-rays) as seen from all viewing angles converge at $t_{\rm obs}\sim 1$ day to $L_{\rm bol}\sim 10^{42}{\rm\, erg\,s^{-1}}$. Meanwhile, the color temperature continues to decrease. To quantify the spectral evolution, we define the color temperature as
\begin{equation}
    T_{\rm c}\equiv\frac{h\nu_{\rm peak}}{2.8k},
\end{equation}
where $k$ is the Boltzmann constant and $\nu_{\rm peak}$ is the peak frequency for spectral luminosity $L_{\nu}$. We also define the peak energy $E_{\rm peak}$ as the energy corresponding to the peak of the $\nu L_{\nu}$ spectrum. The evolution of the color temperature and peak energy for the $Lc$ model is shown in the lower panel of Figure \ref{fig:Lcbol}. The color temperature decreases from $\sim 10^{7}\,$K (keV, X-ray) at $t_{\rm obs}\sim10\,$s to $\sim 10^{4}\,$K (UV/optical) at $\sim1$ day.
}

\begin{figure}
    \centering
    \includegraphics[width=0.9\linewidth]{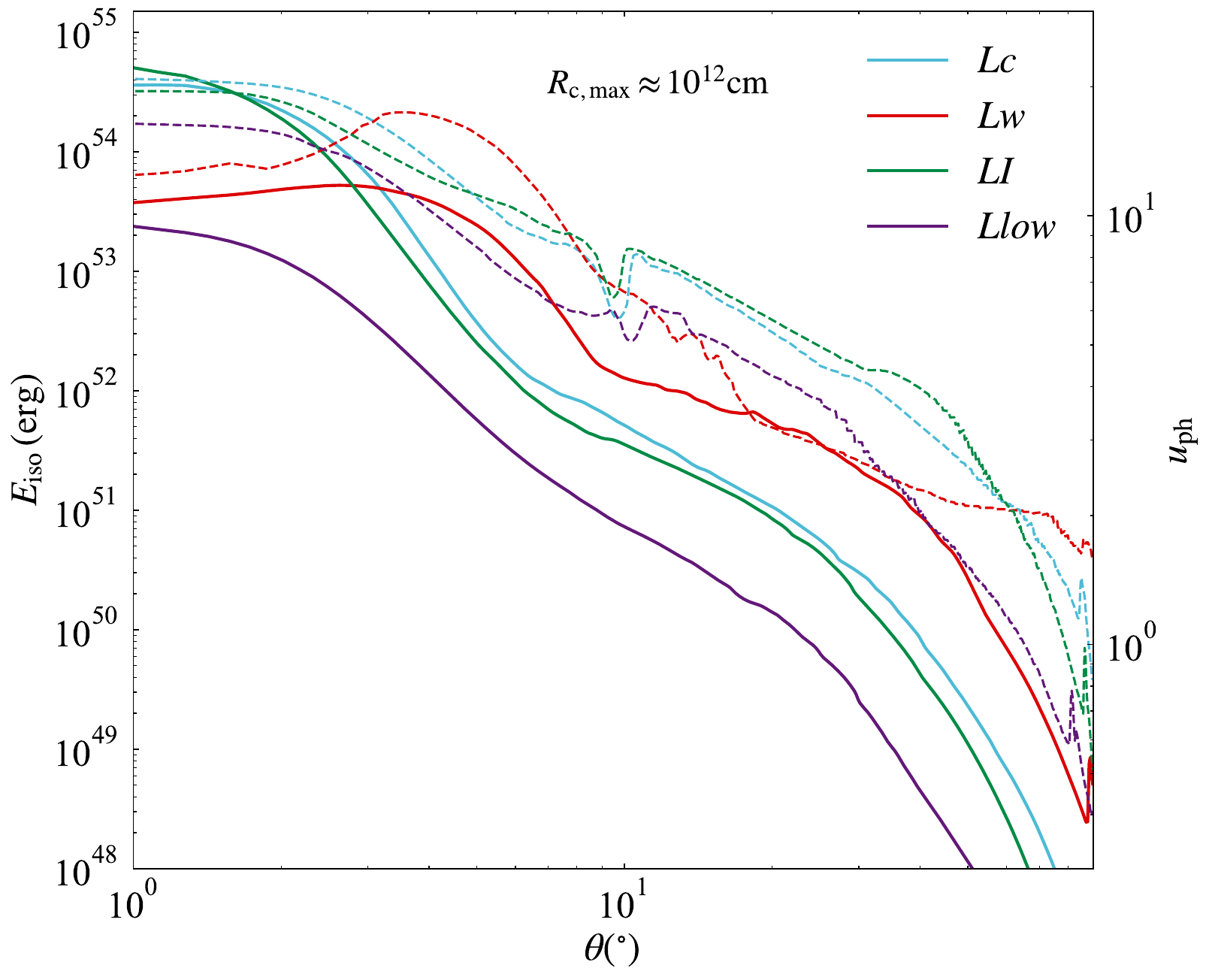}
    \includegraphics[width=0.9\linewidth]{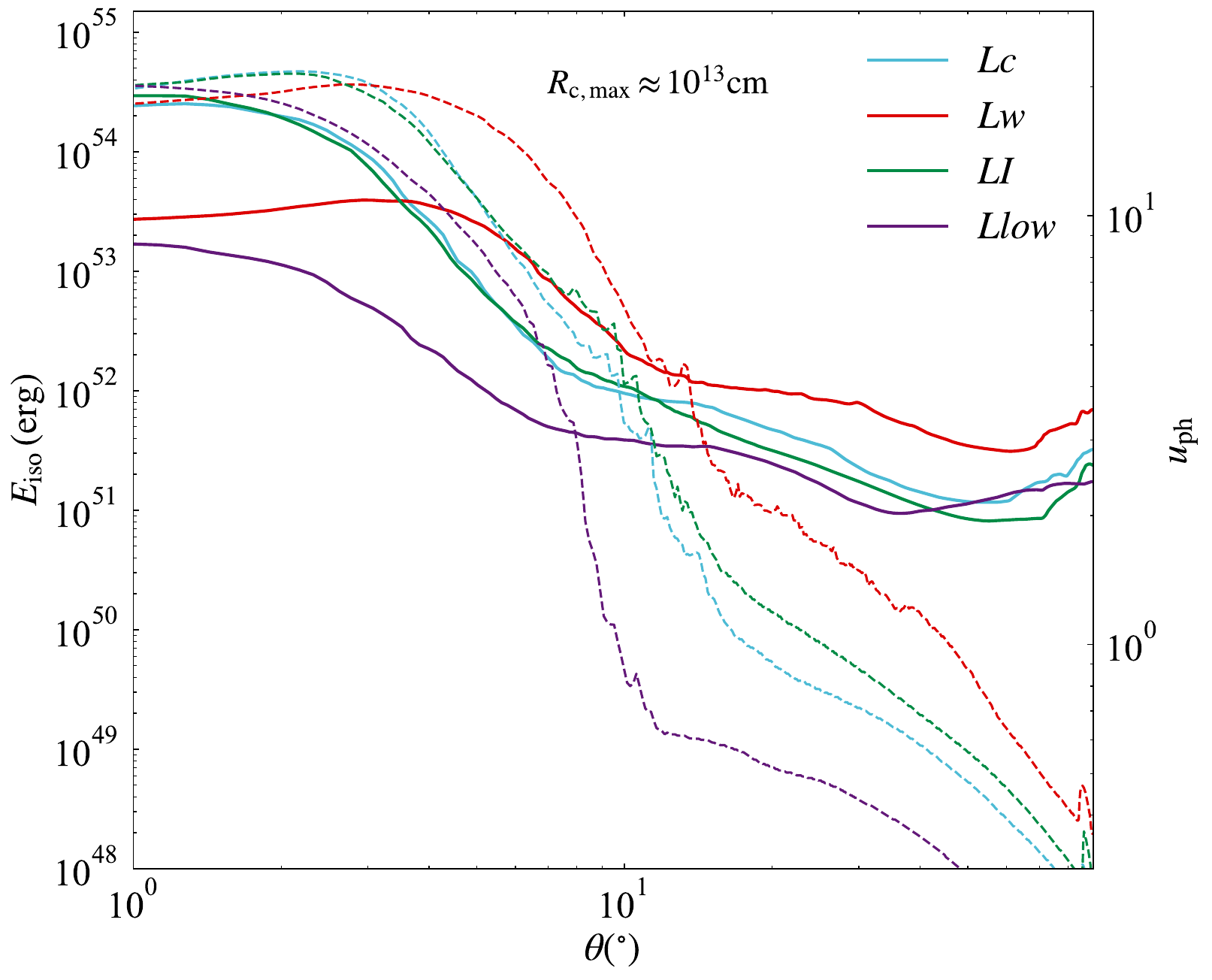}
    \caption{The angular distributions of the isotropic-equivalent energy $E_{\rm iso}$ (solid lines, left axis) and the photospheric four-velocity $u_{\rm ph}$ (dashed lines, right axis) for the four models. The blue, red, green, and purple lines correspond to models $Lc$, $Lw$, $LI$, and $Llow$, respectively. { The upper panel shows the distribution at $R_{\rm c,max}\approx10^{12}\,{\rm cm}$ while the lower panel shows shows those at $R_{\rm c,max}\approx10^{13}\,{\rm cm}$}.
    }
    \label{fig:angular}
\end{figure}

The temporal behavior of non-relativistic cocoon emission has been analytically derived in previous studies by assuming isotropic, homologous expansion. The bolometric lightcurve evolves as $L_{\rm bol}\propto t^{-4/(s+2)}_{\rm obs}$, where the parameter $s$ is determined by the energy distribution in proper velocity $dE/du\propto u^{-s}$ \citep{Nakar2017ApJ...834...28N,Piro2018ApJ...855..103P,Zhu2025MNRAS.544L.139Z}. Previous numerical simulations have found that $s\approx1$ for successful GRB jets \citep[e.g.,][]{Gottlieb2020MNRAS.498.3320G}, implying the decaying slope of the outer cocoon is $L_{\rm bol}\propto t^{-4/3}$. Our simulations also produce a similar energy distribution in four models, leading to a decaying slope of $\sim t^{-1.2}$. The small difference may be attributed to the angular structure of the cocoon.

\begin{figure*}
    \centering
    \includegraphics[width=0.32\textwidth]{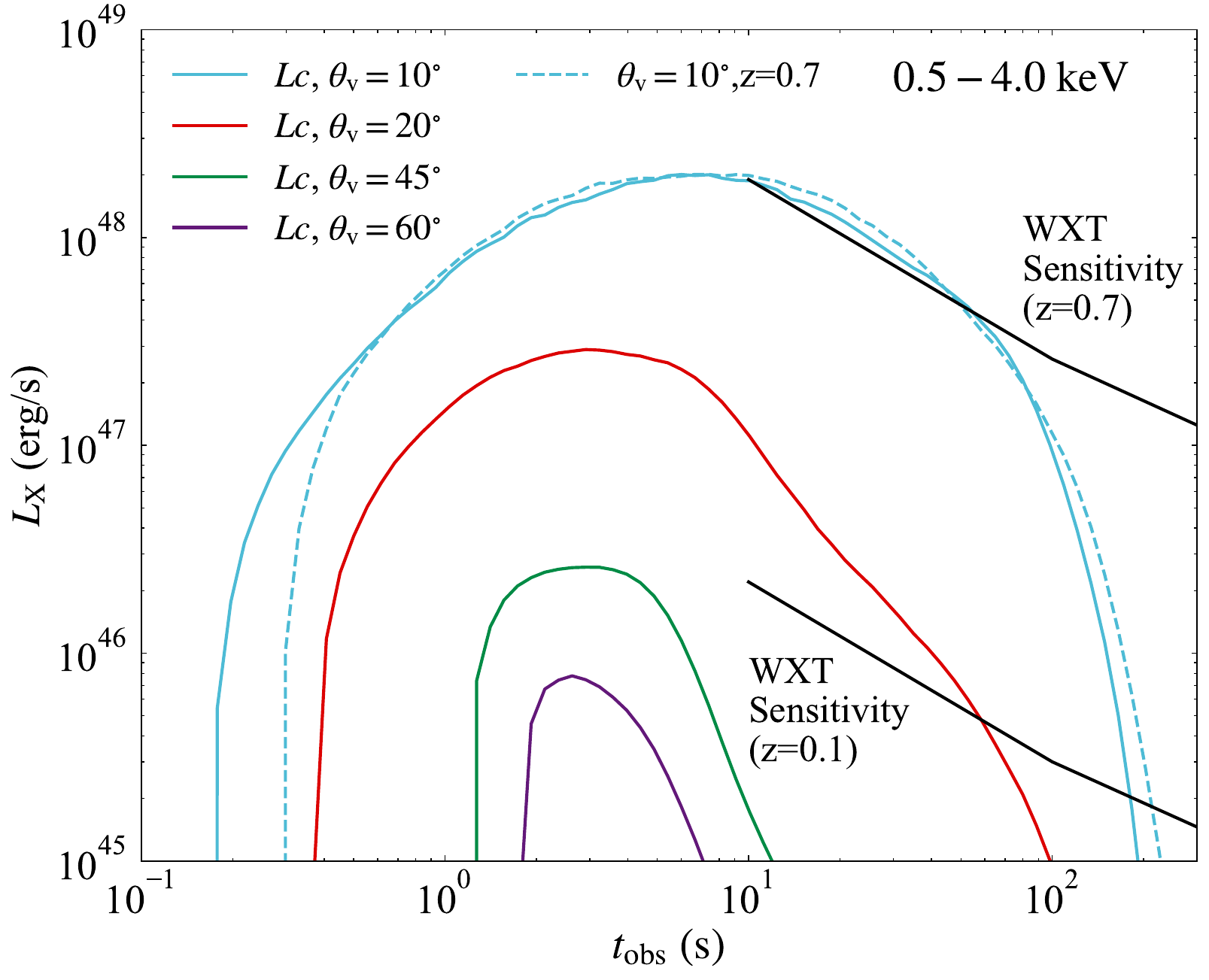}
    \includegraphics[width=0.32\textwidth]{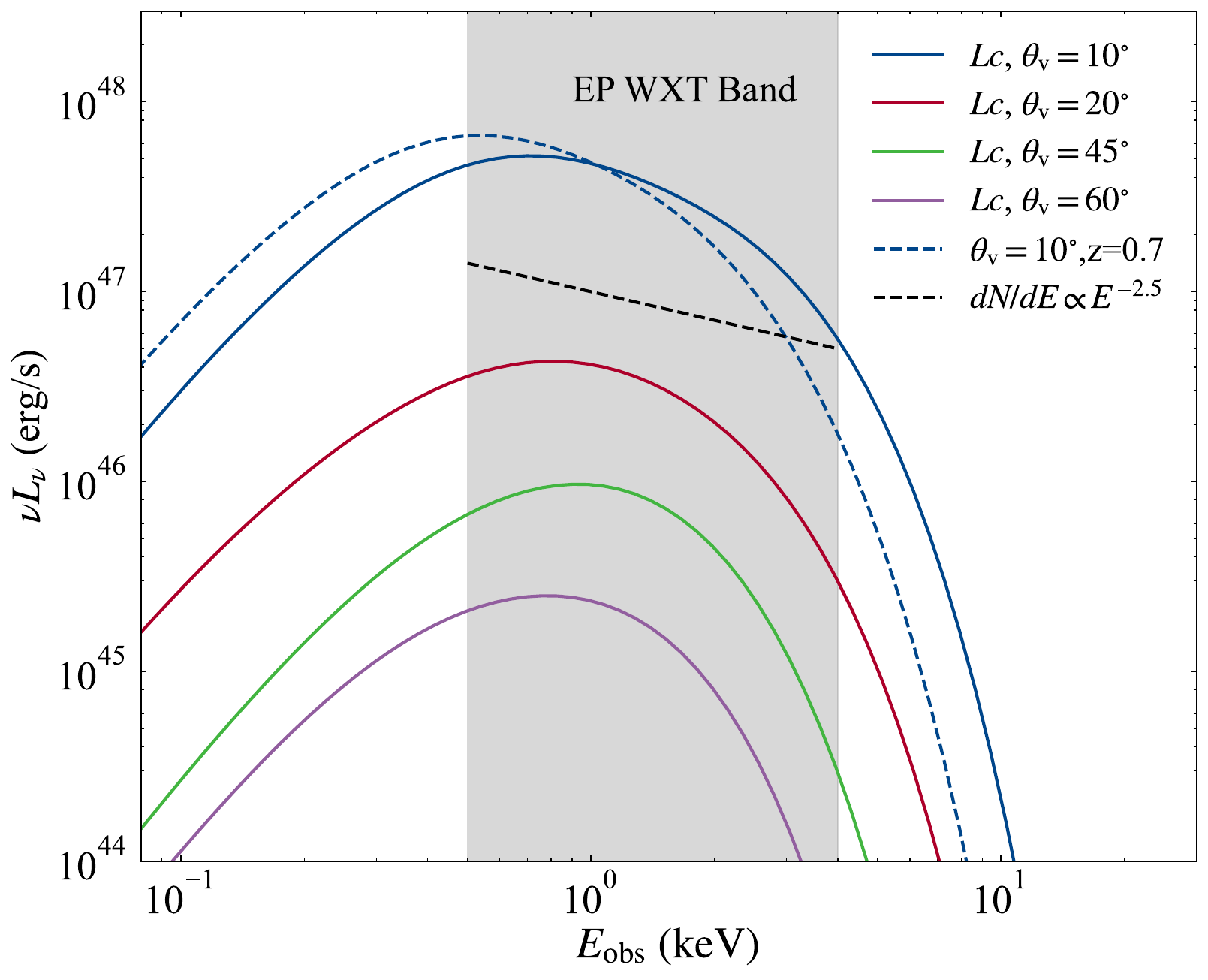}
    \includegraphics[width=0.32\textwidth]{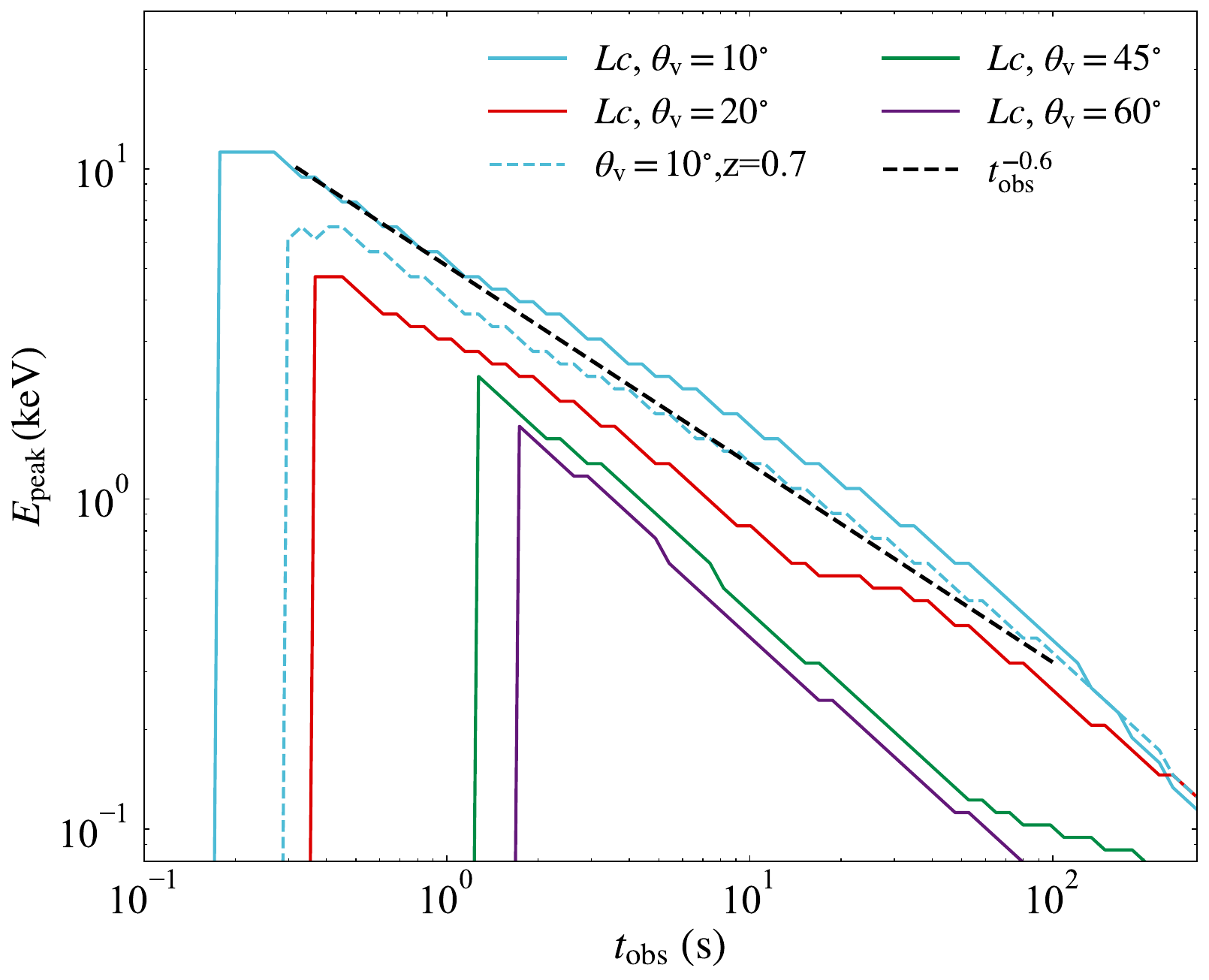}
    \caption{X-ray lightcurves and spectra of the canonical (\textit{Lc}) model. \textit{Left panel}: The X-ray lightcurves in EP WXT band (0.5-4 keV). The blue, red, green, and purple lines correspond to the viewing angles of $10^{\circ}$, $20^{\circ}$, $45^{\circ}$, and $60^{\circ}$. The solid and dashed lines correspond to the X-ray lightcurve produced at redshifts $z=0$ and $z=0.7$. The black solid lines are sensitivity limits of EP WXT  taken from \cite{Yuan2022hxga.book...86Y} for sources at redshifts $z=0.1$ and $z=0.7$.
    \textit{Middle panel}: The $T_{90}$-averaged spectra at different viewing angles.
    The EP WXT band is shown as the gray shaded region. The black dashed line indicates a power-law spectrum with a photon index of 2.5.
    \textit{Right panel}: The evolution of peak energy in \textit{Lc} model.
    The black dashed line indicates a power-law decay rate $E_{\rm peak}\propto t^{-0.6}_{\rm obs}$.}
    \label{fig:Lc}
\end{figure*}

Figure \ref{fig:angular} shows the angular distribution of the $E_{\rm iso}$ and photospheric 4-velocity $u_{\rm ph}=\Gamma_{\rm ph}\beta_{\rm ph}$ when the cocoon front reaches $R_{\rm c,max}\approx10^{12}\,$cm { and $R_{\rm c,max}\approx10^{13}\,$cm}. The isotropic equivalent energy is defined as 
\begin{equation}
    E_{\rm iso}(\theta)=\int^{R_{\rm c,max}}_{R_{\star}}(T_{00}-\Gamma\rho'c^2)4\pi r^2dr,
\end{equation}
where $T_{00}$ is the total energy density. This energy term includes the kinetic energy and thermal energy so it is a conserved quantity for the entire cocoon. 
Different viewing angles probe different values of $E_{\rm iso}$ along the line of sight, so this angular structure also helps explain the viewing-angle dependence of the light curves.
For example, the cocoon has approximately $E_{\rm iso}\approx10^{52}{\rm erg}$ and Lorentz factor $\Gamma_{\rm ph}\approx8$ for $Lc$ model at $\theta_{\rm v}=10^{\circ}$, resulting in a luminosity of $10^{48}{\rm\, erg/s}$ within 100 s, which is consistent with the analytical estimate \citep{Zheng2025ApJ...985...21Z}. { At later times, the angular distribution of $E_{\rm iso}$ becomes flatter at large polar angles due to lateral expansions. As a result, the angular difference in $E_{\rm iso}$ is weakened, leading to a much weaker viewing-angle dependence of the late-time UV/Optical lightcurves.}

{
The difference among the four models arises from the difference in cocoon energy. The cocoon energy $E_{\rm c}\approx L_{\rm j}(t_{\rm bo}-R_{\star}/c)$ is determined by jet luminosity, breakout time, and the stellar radius.

The breakout times are 8\,s, 16\,s, 18\,s, and 13\,s for the canonical (\textit{Lc}), the wide angle (\textit{Lw}), the inflated star (\textit{LI}), and the low energy model (\textit{Llow}). Our breakout times are roughly consistent with the analytical expression calibrated by simulations \citep{Harrison2018MNRAS.477.2128H}
\begin{equation}
    \label{eq:tbo}
    t_{\rm bo}=7{\rm \,s}\, L^{-1/3}_{\rm j,50}R^{2/3}_{\star,11}M^{1/3}_{\star,1}\left(\frac{\theta_{\rm j,0}}{5^{\circ}}\right)^{4/3}.
\end{equation}
The breakout time is mostly sensitive to the initial half opening angle $\theta_{\rm j,0}$, because it significantly affects the opening angle after the collimation $\theta_{\rm j}\propto\theta^2_{\rm j,0}$ \citep{Bromberg2011ApJ...740..100B}, and a smaller open angle pushes the jet head to break out earlier. 
}

{
Although the jet propagating in the inflated star (\textit{LI} case) takes a longer time to break out from the star, the energy deposited into the cocoon $L_{\rm j}(t_{\rm bo}-R_{\star}/c)$ remains comparable to that of the canonical case because of larger $R_{\star}$.
In the \textit{Lw} case, the cocoon energy is roughly a factor of two larger than that in the \textit{Lc}. This is why $E_{\rm iso}$ of $Lc$ and $LI$ are similar and $Lw$ has a more energetic cocoon.
In contrast, the energy of \textit{Llow} case is significantly lower due to its substantially smaller jet luminosity $L_{\rm j}$.
}

\begin{figure}
    \centering
    \includegraphics[width=0.8\linewidth]{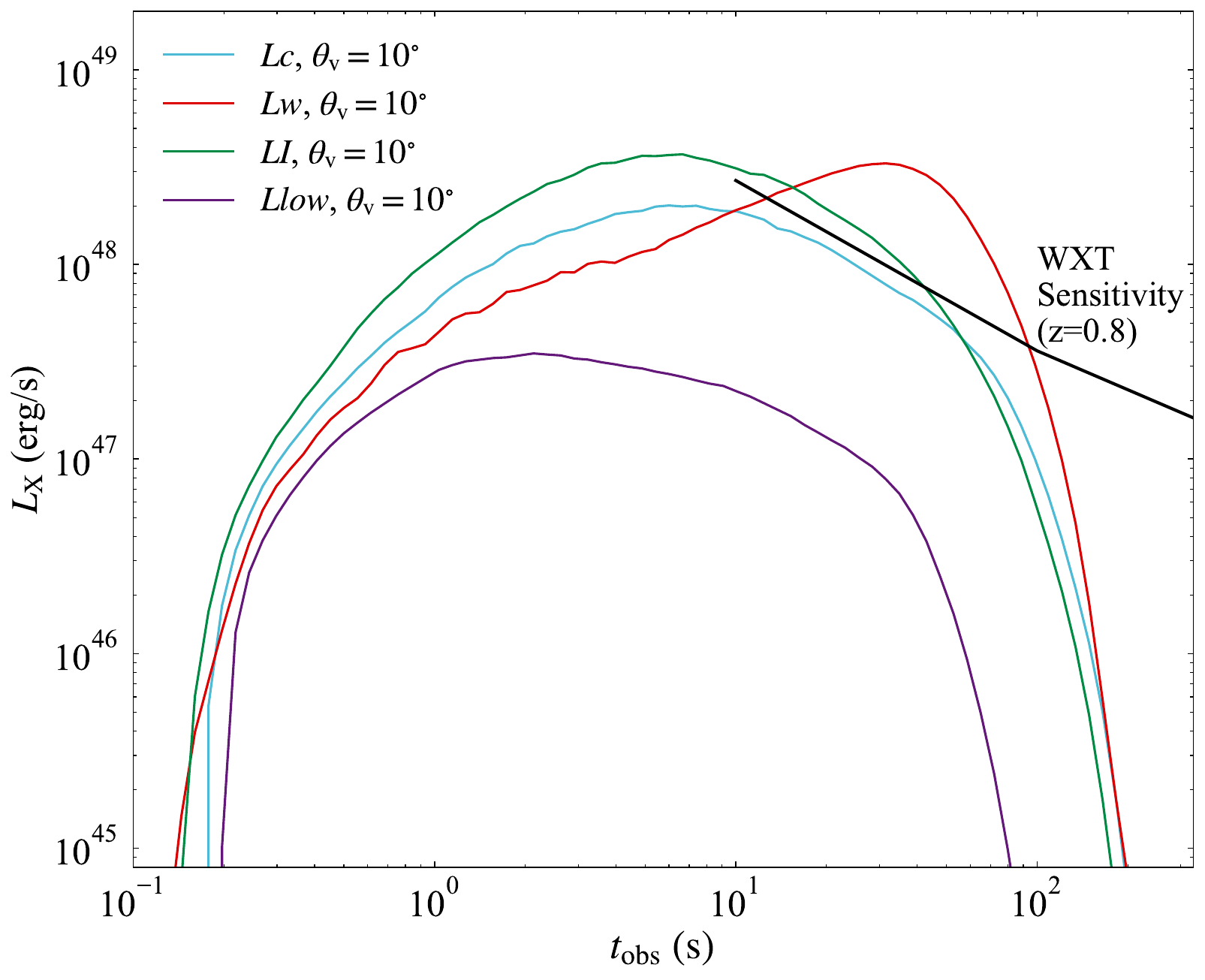}
    \includegraphics[width=0.8\linewidth]{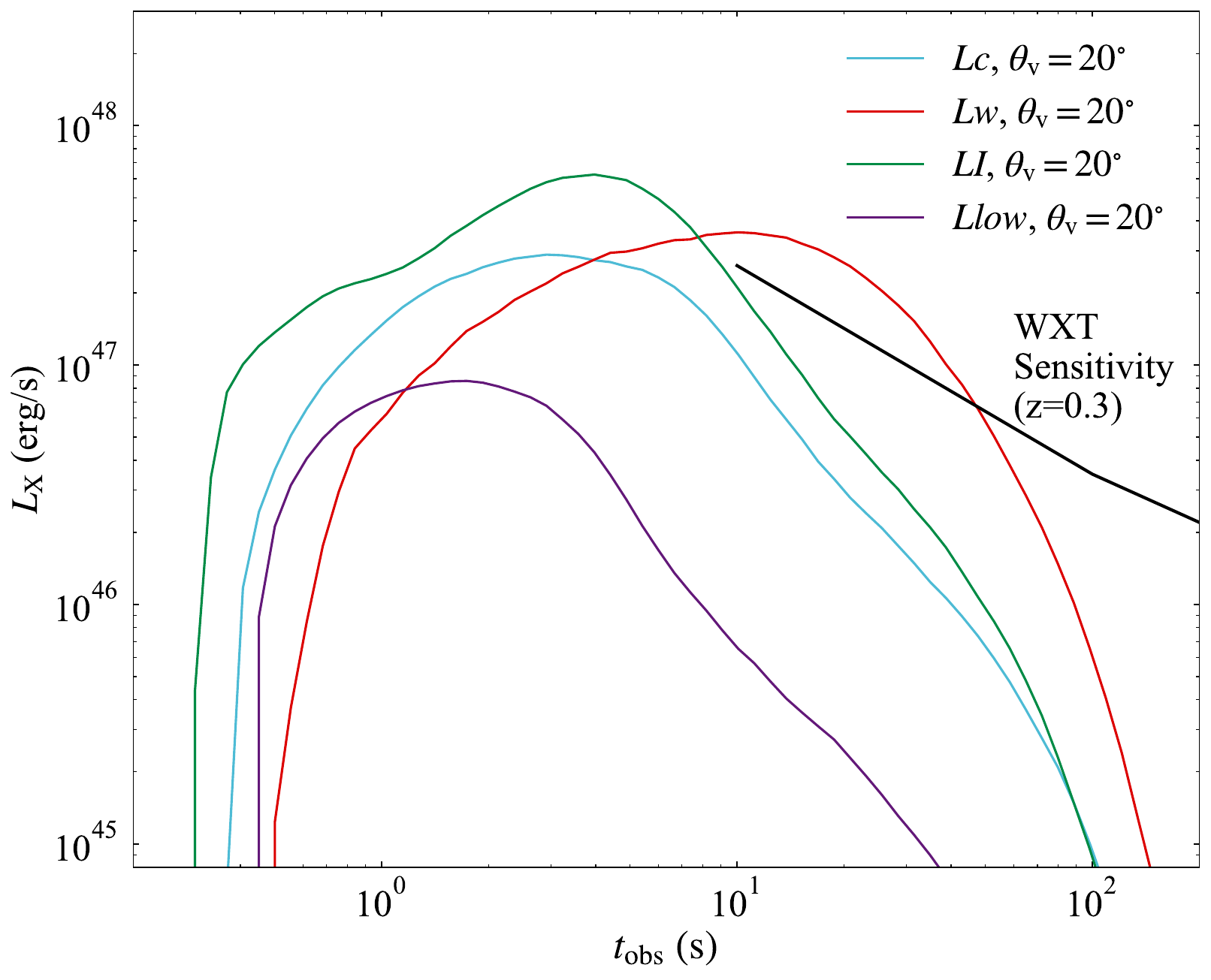}
    \includegraphics[width=0.8\linewidth]{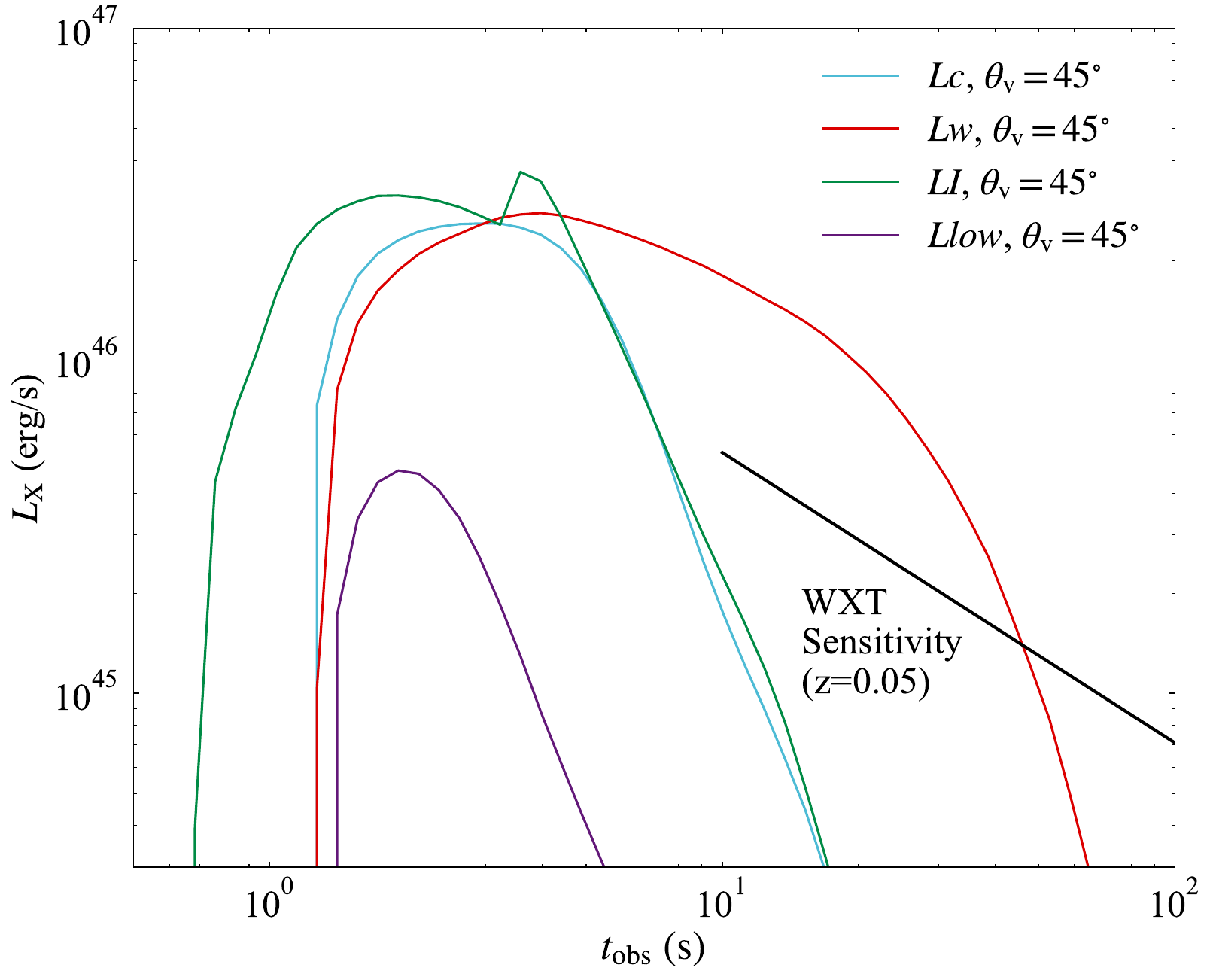}
    \caption{EP WXT (0.5-4 keV) lightcurves in different models. The blue, red, green, and purple lines correspond to model $Lc$, $Lw$, $LI$, and $Llow$. The upper, middle, and lower panels show lightcurves at viewing angles of $\theta_{\rm v}=10^{\circ}$, $20^{\circ}$, and $45^{\circ}$, respectively.
    }
    \label{fig:LXcom}
\end{figure}

\subsection{Fast X-ray Transients}
\subsubsection{General Properties}
Figure \ref{fig:Lc} presents the X-ray lightcurves and spectra as observed at different viewing angles, for our canonical model $Lc$.
We use $L_{\rm X}(t_{\rm obs})$ to denote the lightcurves integrated over the EP WXT band (0.5-4 keV). We find that the X-ray lightcurves rise rapidly within seconds and last for 10-100 s, depending on the viewing angles. The luminosity $L_{\rm X}$ decreases rapidly after the peak because of the rapid cooling of the emitting gas, which causes the peak photon energy to drop below the EP WXT band (see the peak energy evolution on the right panel of Figure \ref{fig:Lc}).

For a slightly off-axis case ($\theta_{\rm v}=10^{\circ}$), the peak luminosity is of the order $10^{48}\,{\rm erg\,s^{-1}}$ and the duration is $\sim\!10^2\,$s, which is detectable for EP WXT up to redshift $z\sim 0.7$. To verify the detectability, we compute the X-ray lightcuvres from a source at $z=0.7$, incorporating all relevant cosmological effects.
At increasing viewing angles, the peak luminosity and duration both decrease because there is less relativistic emitting gas. FXTs seen from large viewing angles ($\theta_{\rm v}\geq45^{\circ}$) are relatively faint ($L_{\rm X}\lesssim10^{46}{\rm erg\,s^{-1}}$) and short-lived ($t_{\rm obs}\lesssim10{\rm\, s}$), which are marginally detectable by WXT in the nearby universe ($z\leq0.05$).

We compute the $T_{90}$-averaged spectra and find that they share a common characteristic: the spectra are generally soft. Here, $T_{90}$ is defined as the time interval between the epochs when 5\% and 95\% of the total energy is released in the WXT band (0.5-4 keV). As shown in the middle panel of Figure \ref{fig:Lc}, the photon index within the WXT band is softer (i.e., greater) than 2.5. This is a distinctive feature that allows us to quickly distinguish the thermal cocoon cooling from the jet's non-thermal emission, whose photon index is generally harder (i.e., smaller) than 2 in the X-ray band. We note that the observed $T_{90}$ is slightly shorter than the theoretical one because the fainter emission is buried by the noise. However, this does not affect the general property of softness.

\begin{figure}
    \centering
    \includegraphics[width=\linewidth]{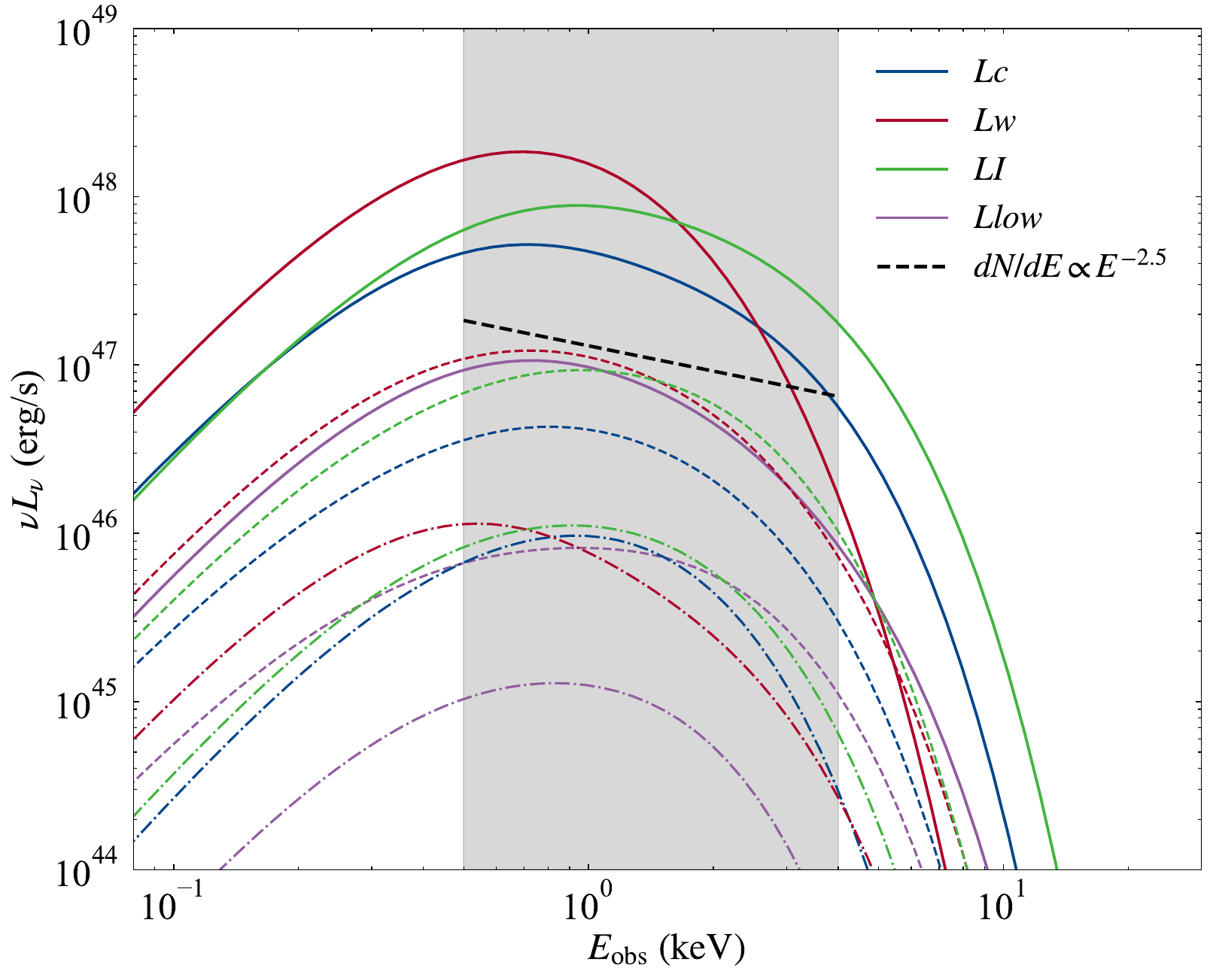}
    \caption{The $T_{90}$-averaged spectra in different models. Colors represent different models, as in Figure \ref{fig:LXcom}. The solid ($\theta_{\rm v}=10^{\circ}$), dashed ($\theta_{\rm v}=20^{\circ}$), and dash-dot ($\theta_{\rm v}=45^{\circ}$) lines represent different viewing angles. The black dashed line is a power-law spectrum with a photon index of 2.5.
    }
    \label{fig:avgmix}
\end{figure}

Another important, general property of the X-ray spectrum is the time evolution of the peak energy, which is shown in the right panel of Figure \ref{fig:Lc}. We find that $E_{\rm peak}$ decreases monotonically with time, following a slope of $\sim t^{-0.6}_{\rm obs}$ which is slightly shallower than the analytical estimate $t^{-11/15}_{\rm obs}$ for the on-axis inner cocoon derived by \cite{Nakar2017ApJ...834...28N}. The shallower evolution may be attributed to contributions from photons emitted at angles away from the LOS that arrive at later times.

\begin{figure}
    \centering
    \includegraphics[width=\linewidth]{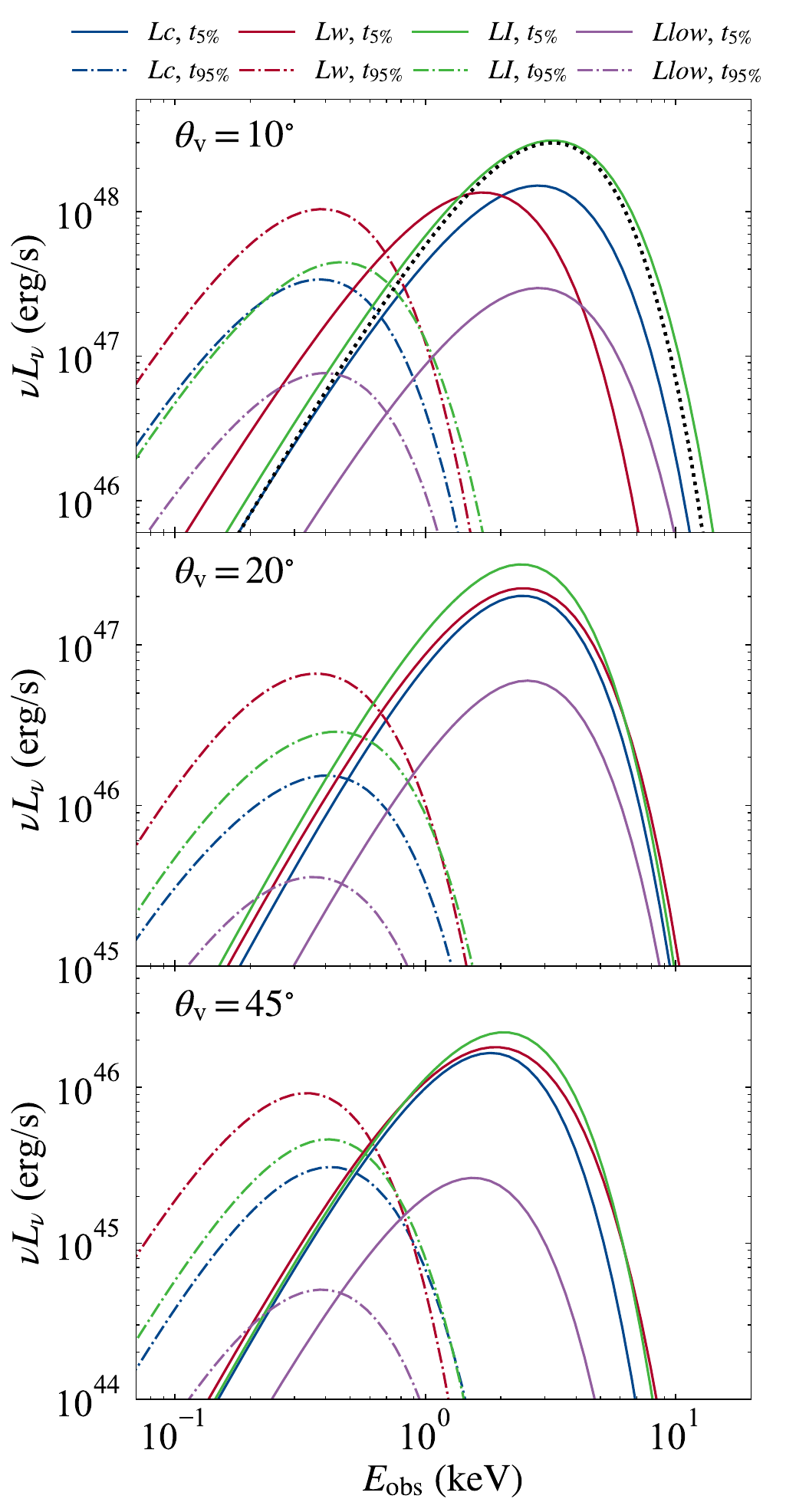}
    \caption{The time-resolved spectra in different models and for different viewing angles. The solid lines are spectra near the beginning of $T_{90}$ ($t_{\rm obs}=t_{5\%}$) and the dashed lines show the spectra near the end of $T_{90}$ ($t_{\rm obs}=t_{95\%}$). Colors represent different models, as in Figure \ref{fig:LXcom}. The black dashed line on the upper panel is the Planck function with $kT=0.8$keV. All spectra are quasi-thermal and only slightly broader than the Planck function. 
    }
    \label{fig:Spec3}
\end{figure}

\subsubsection{Model Comparison}
We compute four models in this work, in all of which the jets successfully break out from the star. Results from these four models are shown in Figures \ref{fig:LXcom} and \ref{fig:avgmix}.
The \textit{LI} case experiences less adiabatic cooling from the stellar surface to the photon diffusion radius. As a result, its emission is expected to be slightly brighter and hotter than that of the \textit{Lc}. In the \textit{Lw} case, the cocoon energy is roughly a factor of two larger than that in the \textit{Lc}, so the emission is also brighter. In contrast, the emission from the \textit{Llow} case is significantly weaker due to its substantially smaller energy budget.

\begin{figure}
    \centering
    \includegraphics[width=\linewidth]{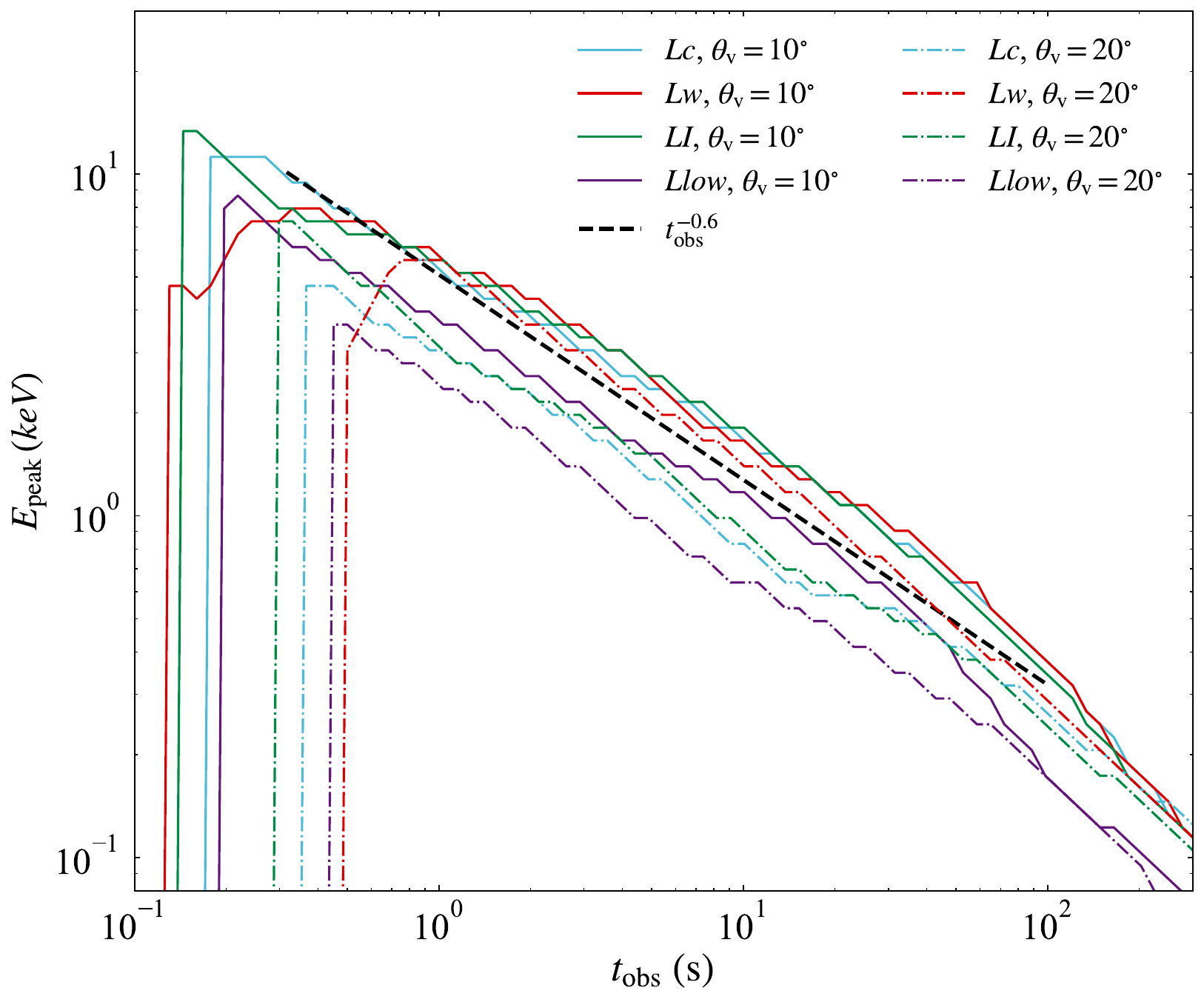}
    \caption{The temporal evolutions of the peak energy in different models. Colors represent different models, as in Figure \ref{fig:LXcom}. Solid lines ($\theta_{\rm v}=10^{\circ}$) and dash-dot lines ($\theta_{\rm v}=20^{\circ}$) indicate different viewing angles. The black dashed line shows the power-law decay $E_{\rm peak}\propto t^{-0.6}_{\rm obs}$.
    }
    \label{fig:Epeak}
\end{figure}


Compared to the canonical case ($\textit{Lc}$), the $T_{90}$-averaged spectra are slightly harder in the \textit{LI} case and softer in the \textit{Lw} case. This behavior arises from the dependence of the observed color temperature on the cocoon energy and stellar radius $T_{\rm c}\propto E^{-1/4}_{\rm c}R^{1/4}_{\star}$ (see eq. 4 in \cite{Zheng2025ApJ...985...21Z}). A larger cocoon energy increases the characteristic diffusion radius $r_{\rm c,diff}\propto E^{1/2}_{\rm c}$ (defined when the total optical depth drops below $\sim \beta^{-1}$) and the corresponding diffusion time $t_{\rm c,diff}\propto E^{1/2}_{\rm c}$ and subsequently reduces the temperature at that radius, leading to a softer spectrum and delayed peak time for the \textit{Lw} case. Following the same scaling relation, the larger stellar radius in the \textit{LI} model leads to a harder spectrum.

The time-resolved spectra presented in Figure \ref{fig:Spec3} reveal more details. We show the spectra at the beginning and the end of the $T_{90}$ interval, corresponding to times when 5\% ($t_{5\%}$) and 95\% ($t_{95\%}$) of the total energy is released (for a given observer). The emission shifts to lower energies during the $T_{90}$ interval for all viewing angles. We find that the instantaneous spectra of these FXTs are quasi-thermal and slightly broader than the Planck function.
The spectral broadening is attributed to two reasons: relativistic motions and the angular temperature profile. Temperature in the observer's frame is boosted by the Doppler factor $T_{\rm obs}=\mathcal{D}T'$, where the Doppler factor depends on the polar angle and the viewing angle. Thus, the spectrum is broadened even for a homogeneous relativistic fireball (see Sec 4.2 of the review by \cite{Peer2017IJMPD..2630018P}). The temperature of the cocoon material generally decreases with polar angle.

Although the \textit{Lw} model is generally softer and \textit{LI} is generally harder in terms of time average spectra, this trend does not necessarily hold throughout the entire $T_{90}$. In the middle of $T_{90}$, the features of time-resolved spectra are consistent with the average spectra because this phase dominates the total emission. However, at the beginning and the end of the $T_{90}$, the instantaneous spectrum deviates from the time-averaged one. For example, as shown in Figure \ref{fig:Spec3},  at $t_{\rm obs}=t_{5\%}$, the \textit{Lw} spectrum for $\theta_{\rm v}=10^{\circ}$ is softer than that from the \textit{Lc} model, whereas it is slightly harder than the \textit{Lc} model for $\theta_{\rm v}=45^{\circ}$.

\begin{figure}
    \centering
    \includegraphics[width=1.0\linewidth]{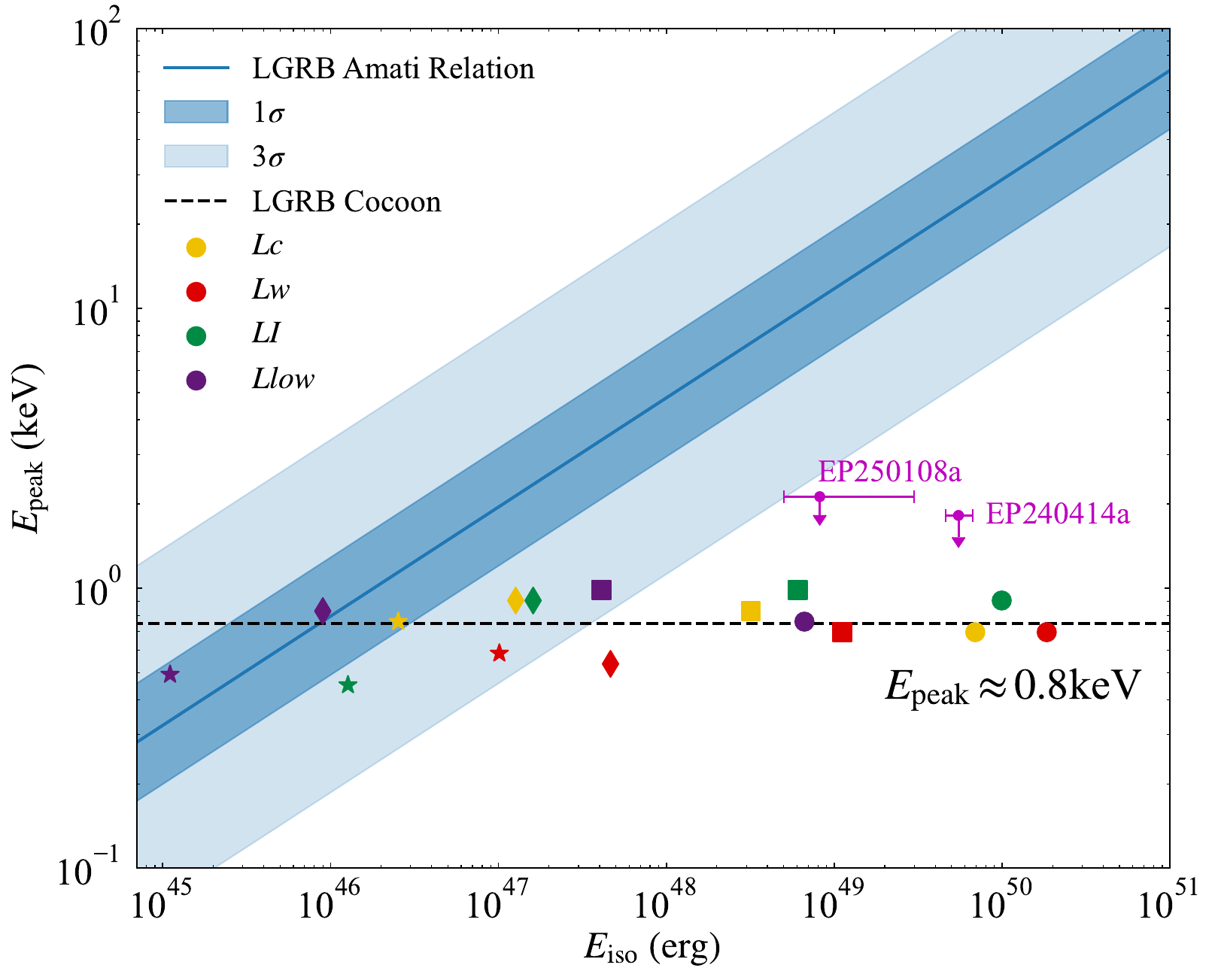}
    \caption{The $E_{\rm iso}$ vs. $E_{\rm peak}$ diagram. The blue solid line represents the Amati relation for the LGRB population, taken from \cite{Liu2025NatAs...9..564L}. The deep blue and light blue shaded regions indicate $1\sigma$  and $3\sigma$ uncertainties, respectively. The yellow ($Lc$), red ($Lw$), green ($LI$), and purple ($Llow$) are results in different models, while the circles ($\theta_{\rm v}=10^{\circ}$), squares ($\theta_{\rm v}=20^{\circ}$), diamonds ($\theta_{\rm v}=45^{\circ}$), and stars ($\theta_{\rm v}=60^{\circ}$) denote different viewing angles. { Magenta points denote observed FXTs (EP240414a and EP250108a), for which $E_{\rm peak}$ only has upper limits.}
    }
    \label{fig:Amati}
\end{figure}

Among the four models, the main observational features are qualitatively similar. Peak luminosities and durations decrease for larger viewing angles. The spectra soften over time during the $T_{90}$ interval and the photon index of the $T_{90}$-averaged spectra is softer than 2.5 across the WXT band. The time evolution of the peak energy in all models is shown in Figure \ref{fig:Epeak}. Overall, $E_{\rm peak}$ decreases roughly as $\sim t^{-0.6}$ regardless of models and viewing angles. A mild steepening occurs after 100\,s, which may originate from the transition from mildly relativistic to trans-relativistic emitting gas.

In Figure \ref{fig:Amati}, we also show the $E_{\rm iso}$ vs. $E_{\rm peak}$ correlation and compare it with the so-called Amati relation inferred from LGRB samples \citep{Amati2002A&A...390...81A}. Here, $E_{\rm peak}$ is the peak energy of the $T_{90}$-averaged spectrum shown in Figure \ref{fig:avgmix}, which is nearly constant $E_{\rm peak}\approx0.8$ keV independent of models and viewing angles. On the other hand, $E_{\rm iso}$ is the isotropic equivalent total radiated energy obtained by integrating the WXT lightcurves $L_{\rm X}(t_{\rm obs})$. For comparison, the Amati relation for LGRBs is $(E_{\rm peak}/70{\rm \, keV})\approx (E_{\rm iso}/10^{51}{\rm \, erg})^{0.4}$ \citep{Liu2025NatAs...9..564L}. FXTs from cocoon cooling emission are a distinct phenomenon compared to the non-thermal gamma-ray prompt emission. This distinction naturally explains why EP continues to discover luminous soft X-ray transients that are clearly separated from the previously known LGRB population.

\subsection{UV/Optical Emission}
\label{sec:opt}
\subsubsection{Early-time UV flash (\texorpdfstring{$t_{\rm obs}\sim$ minutes}{t_obs ~ minutes})}
During the FXT phase, the Rayleigh-Jeans tail ($F_\nu\propto \nu^2$) of the inner cocoon's cooling emission extends to lower frequencies and produces a bright UV flash.
Considering the X-ray luminosity of $L_{\rm X} \sim 10^{47{-}48}{\rm\, erg\,s^{-1}}$ near 0.2 keV (Figure \ref{fig:avgmix}) for slightly off-axis viewing angles, the luminosity near 10 eV { is predicted to be} $L_{\rm UV} \sim 10^{43-44}{\rm\, erg\,s^{-1}}$ in the first few minutes in the observer frame. 

\begin{figure}
    \centering
    \includegraphics[width=1.0\linewidth]{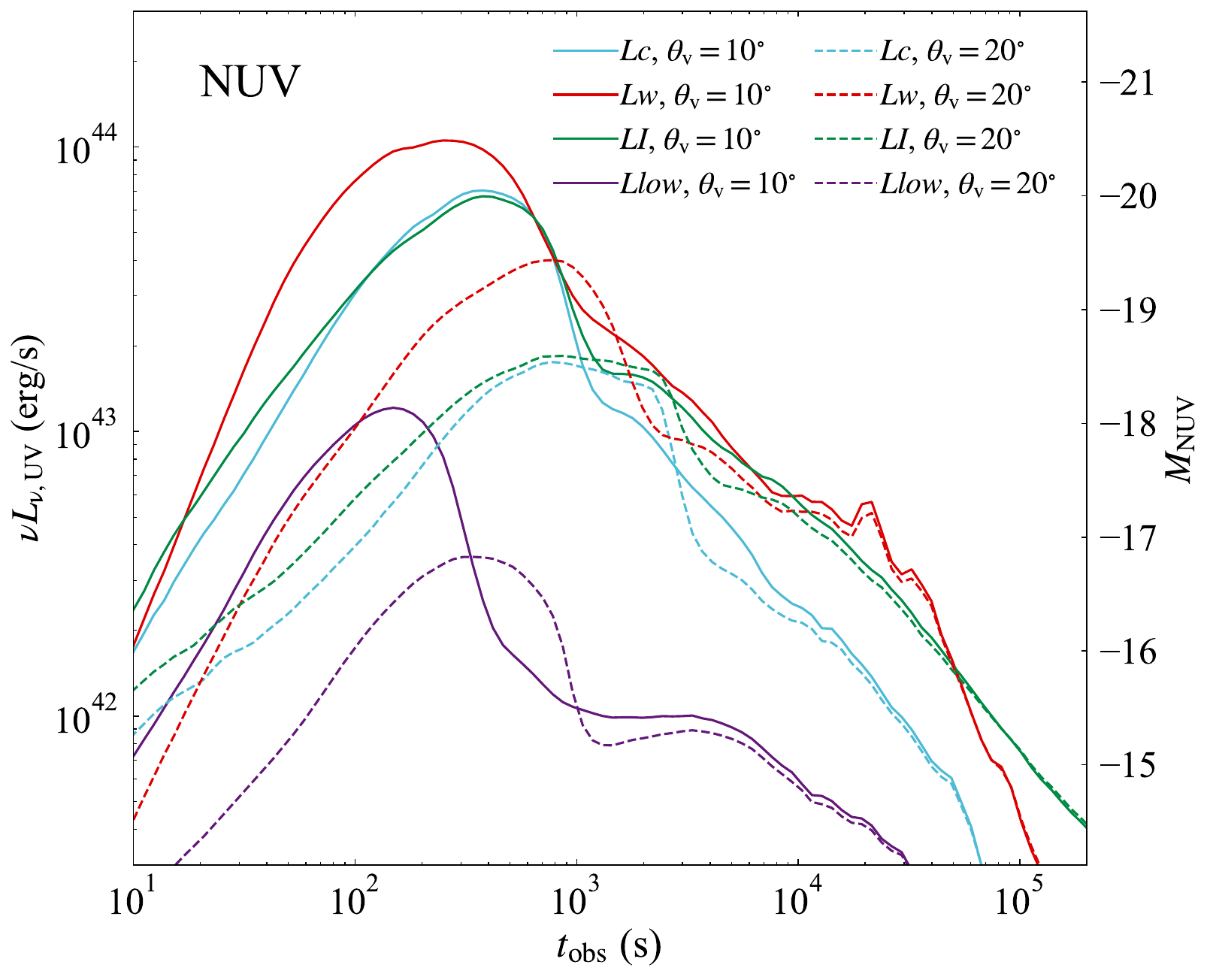}
    \caption{Lightcurves at observer's wavelength $\lambda= 200\,$nm ($\approx$NUV band). Colors represent different models, as in Figure \ref{fig:LXcom}. Solid lines ($\theta_{\rm v}=10^{\circ}$) and dashed lines ($\theta_{\rm v}=20^{\circ}$) indicate different viewing angles. The left axis shows the luminosity, while the right axis gives the corresponding absolute magnitude.
    }
    \label{fig:UV}
\end{figure}

We show the lightcurves in the Near Ultraviolet (NUV) band (=NVW2 band on Swift UVOT) in Figure \ref{fig:UV}. We find that the early UV flash lasts for a few hundred seconds, corresponding to the FXT duration (but slightly longer). The luminosity of the UV flash is boosted by relativistic motion so it is only apparent for $\theta_{\rm v}\lesssim20^{\circ}$. At large viewing angles, the UV flash is very weak, and the lightcurves closely follow a power-law decay (see the black dashed line in Figure \ref{fig:Lcbol}).
The space-based UV telescopes \citep[e.g., Swift UVOT or UVEX,][]{Gehrels2004ApJ...611.1005G,Singer2025PASP..137g4501S} may detect the UV flash if they can rapidly follow up the EP trigger. The Ultraviolet Transient Astronomy Satellite (ULTRASAT) could potentially detect the UV flash independently \citep{ULTRASAT}. Detection of the predicted early-time UV flash (especially coincident with FXT) will confirm the thermal nature (or cocoon origin) of the X-ray emission.

\subsubsection{Late-time UV decay and optical plateau (\texorpdfstring{$t_{\rm obs}\sim$ hours to days}{t_obs ~ hours to days})}

After the photons in the relativistic inner regions of the cocoon have diffused out, the trans-relativistic and non-relativistic regions at larger polar angles dominate the cooling emission. { Weaker beaming effect leads to quasi-isotropic radiation and allows the observed luminosity to be determined by a few global cocoon parameters.}

\begin{figure}
    \centering
    \includegraphics[width=0.95\linewidth]{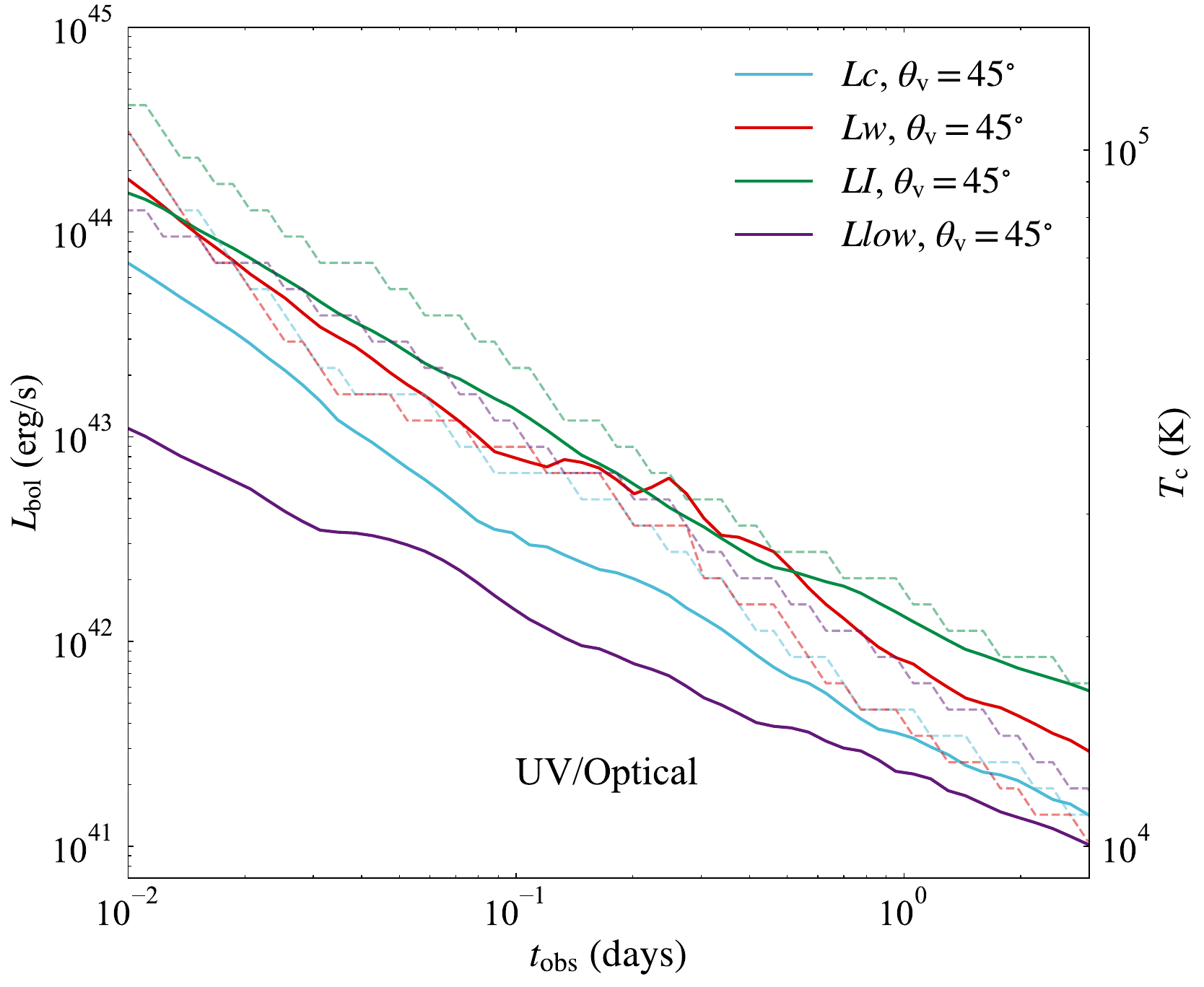}
    \caption{Late-time bolometric lightcurves (left axis) for different models viewed at $\theta_{\rm v}=45^\circ$. Colors represent different models, as in Figure \ref{fig:Lc}. The dashed lines are the corresponding color temperatures (right axis).}
    \label{fig:Lbolmix}
\end{figure}

The luminosity of the cocoon emission can be estimated by accounting for the adiabatic loss during propagation \citep{Nakar2017ApJ...834...28N,Zheng2025ApJ...985...21Z}. Assuming the cocoon has initial thermal energy $E_{\rm c,th0}$ and volume $V_{\rm c0}$ when it breaks out, the remaining thermal energy after expanding to a volume $V_{\rm c}$ is $E_{\rm c,th}=E_{\rm c,th0}(V_{\rm c}/V_{\rm c0})^{-1/3}$. The volume change includes the radial and lateral expansions $V_{\rm c} \propto (1-\cos\theta_{\rm c})R^3_{\rm c}$, where $\theta_{\rm c}$ is the half opening angle of the cocoon. Near the breakout ($R_{\rm c}\sim R_\star$), the initial cocoon opening angle $\theta_{c0}$ is small such that $1-\cos\theta_{\rm c0}\approx \theta^2_{\rm c0}/2$. 
Quickly after the breakout, the cocoon expands laterally to a quasi-spherical configuration with $\theta_{\rm c}\sim 1\rm\, rad$. Therefore, the volume expansion factor is given by $V_{\rm c}/V_{\rm c0}\simeq 2(R_{\rm c}/R_\star)^3/\theta^2_{\rm c0}$. As we show in Figure \ref{fig:Engden}, most photons diffuse out when the cocoon expanded to the \textit{global} diffusion radius of $R_{\rm c,diff}\simeq 10^{14}\rm\, cm$, which takes a diffusion time of $t_{\rm c,diff}\simeq 1\,$day. Therefore, the bolometric luminosity at $t_{\rm obs}=t_{\rm c,diff}\simeq 1\rm\, day$ can be estimated by

\begin{equation}
\label{eq:Lbol}
\begin{aligned}
    L_{\rm bol}&(t_{\rm c,diff}\simeq 1\,{\rm day})
    \simeq \frac{E_{\rm c,th}(R_{\rm c,diff})}{t_{\rm c,diff}} \\
    &  \simeq 10^{42}\,{\rm erg\,s^{-1}}\, E_{\rm c,th0,50.5} R_{\star,11} \left(\frac{\theta_{\rm c0}}{10^{\circ}}\right)^{2/3}.
\end{aligned}
\end{equation}

Our numerical results are consistent with the analytical estimate. Eq. \ref{eq:Lbol} also indicates that the \textit{LI} and \textit{Lw} models should be brighter than the \textit{Lc} model around 1 day, which is also confirmed by the numerical results shown in Figure \ref{fig:Lbolmix}. The luminosity ratio between \textit{LI} and \textit{Lc} is approximately 4, in excellent agreement with the analytical scaling.

\begin{figure}
    \centering
    \includegraphics[width=0.9\linewidth]{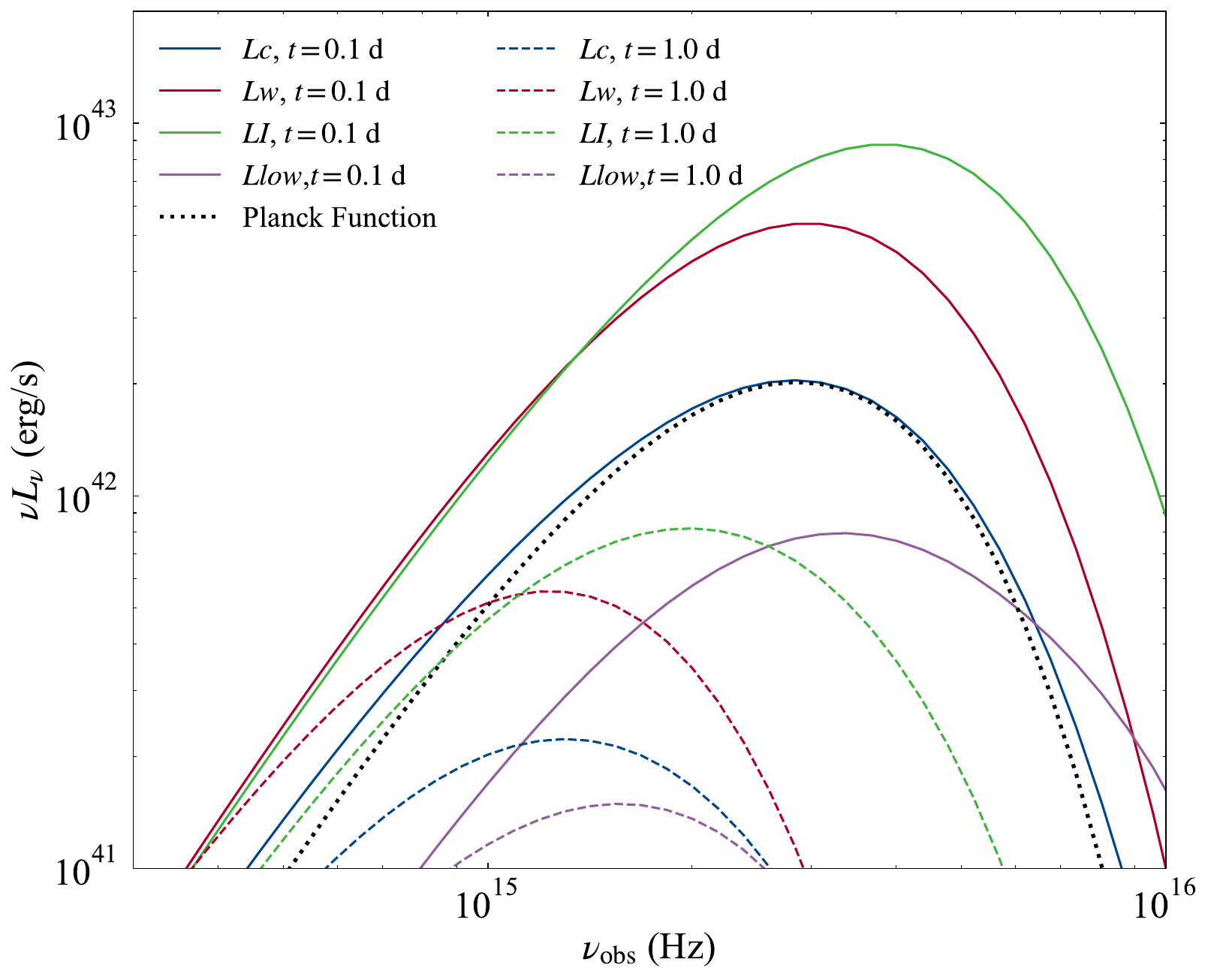}
    \caption{Time-resolved spectra at late times. The solid lines are spectra at 0.1 days, and the dashed lines show the spectra at 1 day. Colors represent different models, as in Figure \ref{fig:LXcom}. The black dashed line is the Planck function with $T=35000$ K. The observed spectra are quasi-thermal but slightly broader than the Planck function. All models shown are for $\theta_{\rm v}=45^{\circ}$.
    }
    \label{fig:UVspec}
\end{figure}

The time-resolved UV/optical spectra are shown in Figure \ref{fig:UVspec}. Similar to the X-ray emission from the inner cocoon, the UV/optical emission exhibits quasi-thermal spectra that are slightly broader than the Planck function. The peak frequency of the luminosity is located in the UV band in all models, indicating that optical telescopes should detect a blue transient.

The color temperature evolutions are presented in Figure \ref{fig:Lcbol}$\&$\ref{fig:Lbolmix}. The color temperatures in all models monotonically decay as a power-law with a slope $\sim t^{-0.33}_{\rm obs}$, which is consistent with the analytical estimate $T_{\rm c}\propto t^{-0.38}_{\rm obs}$ in \cite{Nakar2017ApJ...834...28N}. The temperature in the \textit{LI} model is always slightly higher than in other models at a given observer's time, as there is less adiabatic loss after breaking out from a larger stellar radius. The color temperatures in all models are still comparable to each other because they are not very sensitive to the cocoon energy ($T_{\rm c}\propto E^{-1/4}_{\rm c}$) and other parameters \citep{Zheng2025ApJ...985...21Z}.

\begin{figure}
    \centering
    \includegraphics[width=\linewidth]{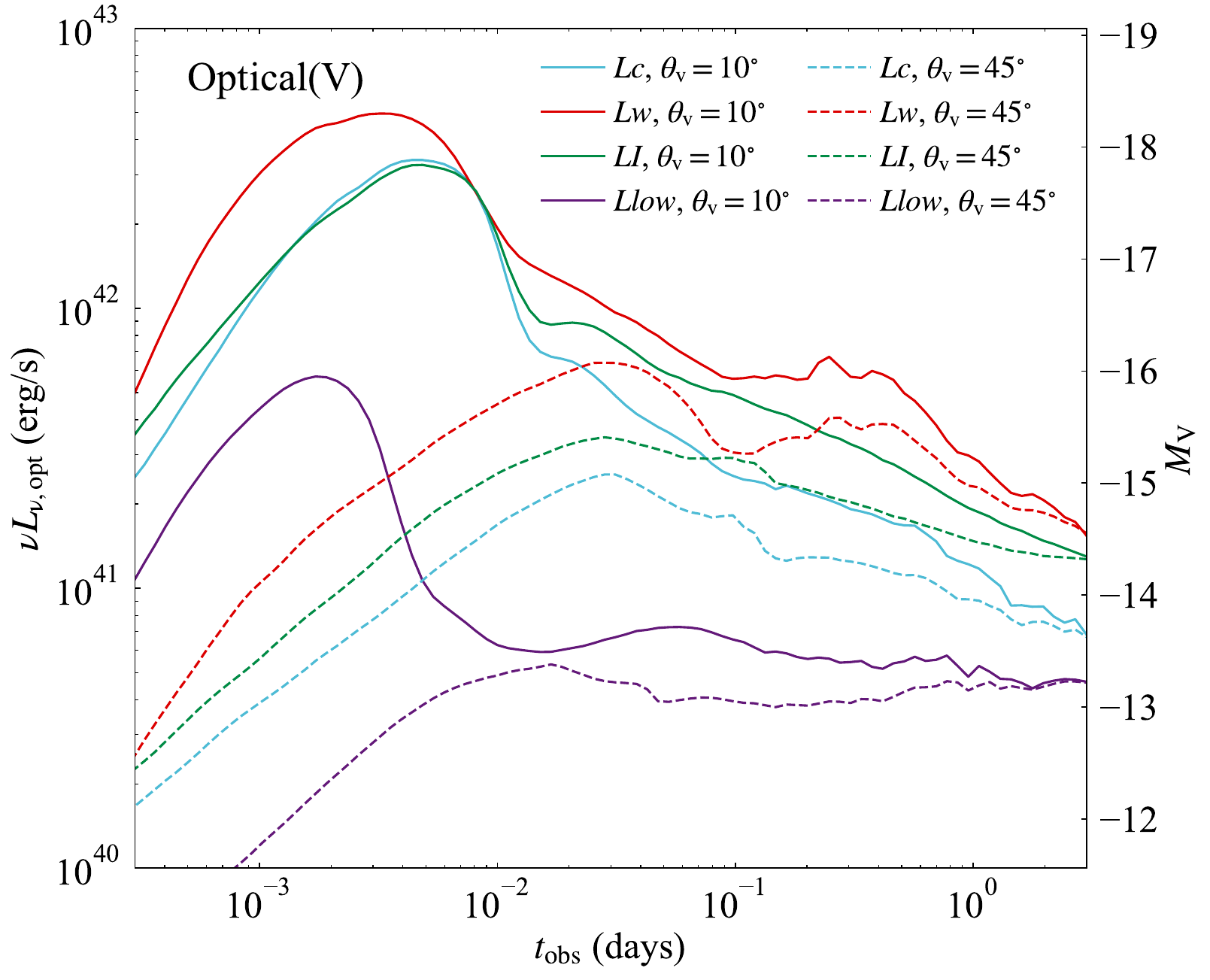}
    \caption{Optical lightcurves at $\lambda=555$ nm ($\approx$V band).  Solid lines ($\theta_{\rm v}=10^{\circ}$) and dashed lines ($\theta_{\rm v}=20^{\circ}$) are for different viewing angles. Colors represent different models, as in Figure \ref{fig:LXcom}.}
    \label{fig:Lopt}
\end{figure}

For $T_{\rm c}\gtrsim10^{4}\,$K, optical telescopes only observe the Rayleigh-Jeans tail of the cooling emission. Consequently, the optical luminosity ($\nu L_\nu$) is expected to be on the order of $10^{41}{\rm erg\,s^{-1}}$ at $t_{\rm obs}\simeq 1\rm\,$day.  The optical lightcurves in different models are shown in Figure \ref{fig:Lopt}.
The optical lightcurve decay is flatter than that of the bolometric lightcurve, because the observed band is below the spectral peak. The temporal slope of optical lightcurve decay is given by $L_{\rm opt}\propto L_{\rm bol}/T^3_{\rm c}\propto t^{-0.2}_{\rm obs}$. Thus, we expect to see an \textit{optical plateau}, which is qualitatively similar to those observed in some FXTs \citep[e.g., EP 240414a, EP 250108a][]{vanDalen2025ApJ...982L..47V,Rastinejad2025ApJ...988L..13R}. The optical plateaus from cocoon cooling are thermal and blue, in contrast to the non-thermal red plateaus expected from GRB afterglow emission (including the afterglow produced by the cocoon-stellar wind interactions). These two components can be easily distinguished through their photometric color and temporal evolution. 

{ We note that the opacity is assumed to be constant in both the analytical and numerical calculations. This approximation may break down at late times, when the temperature drops below $T_{\rm c} \lesssim 10^{4}\,$K and the opacity decreases (leading to slightly higher $L_{\rm bol}$).
}

Although our cocoon cooling model successfully explains the timescale, color, and the lightcurve shape of the early UV/optical emission observed in some FXTs, the predicted luminosity is unfortunately much fainter than observations. Several FXTs have very bright UV/optical emission ($L_{\rm bol}\gtrsim 10^{43}{\rm erg\,s^{-1}}$) at $t_{\rm obs}\simeq 1\,$day \citep[EP 240414a, EP 250108a,][]{vanDalen2025ApJ...982L..47V,Rastinejad2025ApJ...988L..13R}, which are at least one order of magnitude brighter than our results. Solving this discrepancy is beyond the scope of the current work because we focus on the FXT phase, for which we successfully reproduce the observed X-ray luminosity, duration, and spectrum. Some possible solutions to the UV/optical luminosity discrepancy are discussed in \S\ref{sec:lumino}.
Those solutions do not affect the spectral properties of FXTs because the early-time X-ray emission comes from distinct emitting regions (the inner cocoon). Notably, the FXT timescales are weakly affected because they depend primarily on the Lorentz factor of the gas along the LOS $t_{\rm c,diff}\propto E^{1/2}_{\rm c}\Gamma^{-5/2}_{\rm c}$, and we find $\Gamma_{\rm c}\sim\,$a few in all models that have a successful jet breakout.


\section{Discussion}\label{sec:dis}

In this section, we discuss the detection rate, some unsolved issues, and potential extensions of our model.

\subsection{Detection rate}
{ The WXT detection rate of off-axis cocoons depends on the rate of SNe Ic-BL and their jet fraction. Since SNe Ic-BL are only a small portion ($\sim3\%$) of core-collapse supernovae (D. Perley et al. 2026, in preparation), the local event rate density is $\mathcal{R}_{\rm Ic\mbox{-}BL}\sim3\times10^{3}\,{\rm Gpc}^{-3}{\rm yr}^{-1}$. The jet fraction is estimated to be $1\%\lesssim f_{\rm jet}\lesssim 10\%$ based on FXT detection and radio follow-ups for SNe Ic-BL \citep{Zheng2025ApJ...985...21Z}.  
EP covers $\sim\!8.7\%$ of the sky with a duty cycle of $50\%$ \citep{Yuan2022hxga.book...86Y,Sun2025NatAs...9.1073S}. For the brightest cocoon (e.g. $\theta_{\rm v}\simeq 10^{\circ}$) that can be detected at $z=0.7-0.8$, the beaming factor is $f_{\rm b}\approx0.015$. Combining all these factors, we estimate the detection rate of the brightest cocoon emission to be $\sim 10 f_{\rm jet,-1}{\rm yr^{-1}}$. 
Because the detection volume increases rapidly with luminosity, the observed sample is dominated by the brightest events. As a result, the total detection rate is expected to be comparable to that of the brightest population.
}


\subsection{Non-thermal prompt emission from within jets}

For a successful jet breakout, we expect the ultra-relativistic jet core to produce non-thermal prompt gamma-ray emission that is strongly beamed near the jet axis. It is interesting to compare the prompt gamma-ray emission viewed from off-axis with the thermal cocoon emission.

\begin{figure}
    \centering
    \includegraphics[width=0.9\linewidth]{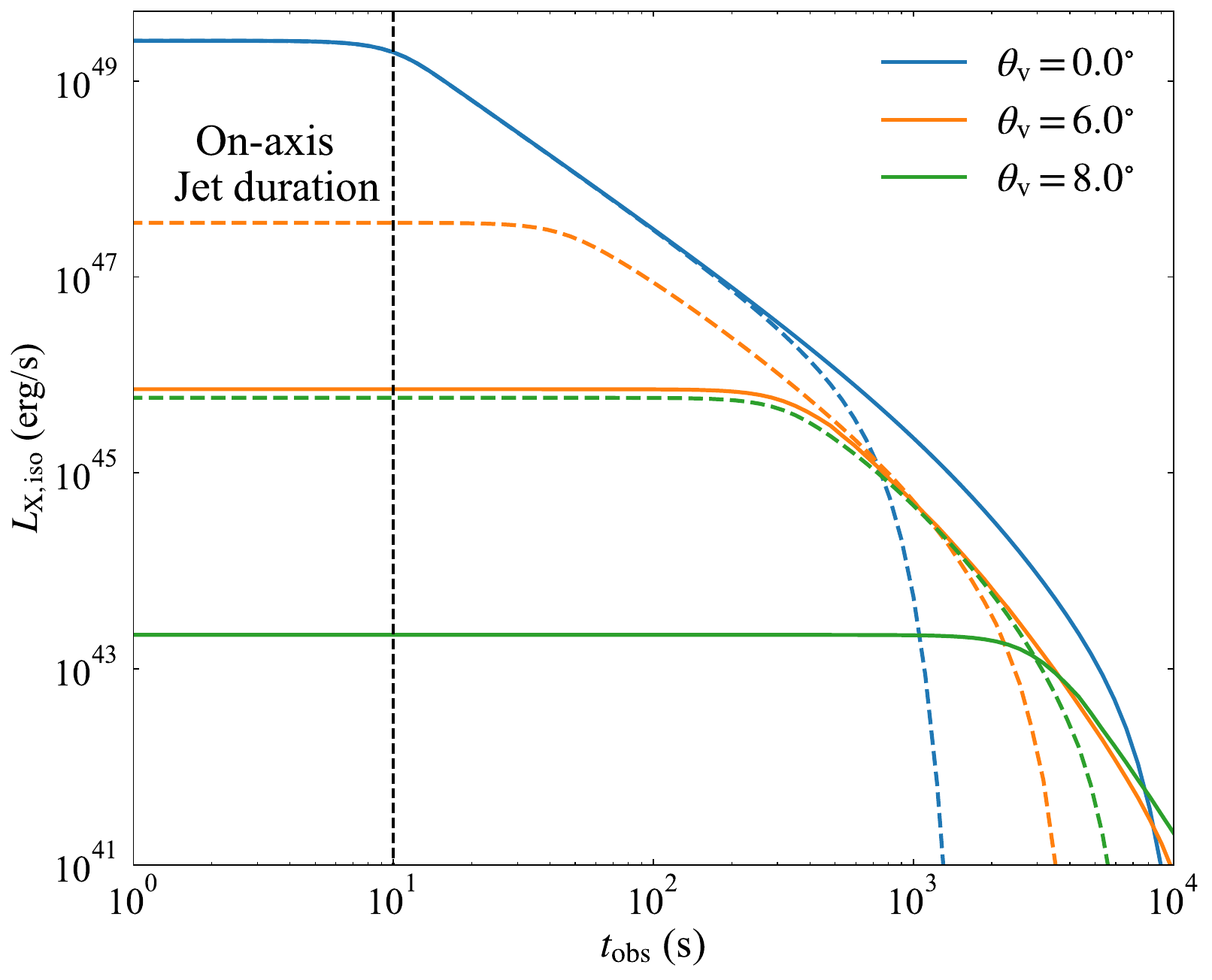}
    \caption{Non-thermal X-ray ``prompt emission'' at $E_{\rm obs}=1\,$keV in the observer's frame, based on the extrapolated Band function. The jet duration is $t_{\rm obs}=10\,$s for on-axis ($\theta_{\rm v}=0^{\circ}$) observers. In the on-axis lightcurves, the power-law decay at $t_{\rm obs}>10\,$s arises from the high-latitude emission. The blue ($\theta_{\rm v}=0^{\circ}$), orange ($\theta_{\rm v}=6^{\circ}$) and green ($\theta_{\rm v}=8^{\circ}$) lightcurves correspond to different viewing angles. Solid and dashed lines are for different jet Lorentz factors of $\Gamma_{\rm j}=$300 and 100, respectively.  
    }
    \label{fig:prompt}
\end{figure}

An interesting regime is when the observer's LOS is located near the edge of the ultra-relativistic jet core. For instance, at polar angles $5^{\circ}\lesssim\theta\lesssim10^{\circ}$, the jet Lorentz factor is may be lower than that of the core, $10\lesssim\Gamma\lesssim100$. In this regime,
a mixture of thermal and non-thermal emissions may emerge because the jet-cocoon system is transitioning from the non-thermal prompt emission to the thermal cocoon emission. For instance, the X-ray spectrum may show a thermal bump above the underlying non-thermal power-law continuum. 
The difference between non-thermal and thermal components also appears in their lightcurves. Thermal emission cuts off when the peak energy drops below the observer's band, while the non-thermal emission can decay as a power-law due to the contribution from high-latitude emissions \citep[e.g.,][]{Kumar2000ApJ...541L..51K}. As a result, we may expect to observe a power-law tail in the X-ray lightcurve that is not captured by our thermal cocoon emission model (see Figure \ref{fig:prompt}). In the power-law tail, the X-ray spectrum will harden with time following the closure relation $F_{\rm\nu, obs}\propto\nu^{\beta}_{\rm obs}t^{2-\beta}_{\rm obs}$. Afterwards, the off-axis afterglow emission arrives at the observer, likely producing an X-ray plateau \citep{2020MNRAS.492.2847B}.

We compute the high-latitude emission and present the results in Figure \ref{fig:prompt}. We assume the jet has a top-hat structure with half opening angle of $5^{\circ}$ and releases isotropic gamma-ray energy of $E_{\rm\gamma,iso}=10^{53}{\rm \,erg}$ within $t_{\rm obs}=10\,$s for on-axis observers, corresponding to typical LGRBs in Fermi GBM samples \citep{Poolakkil2021ApJ...913...60P}. The prompt gamma-ray spectrum is modeled by the Band function with an observed peak energy $E_{\rm peak}=200\,$keV, low energy photon index $-0.84$ and high energy photon index $-2.4$, which are also typical values for LGRBs \citep{Poolakkil2021ApJ...913...60P}. We adopt two cases for the bulk Lorentz factor of the jet as $\Gamma_{\rm j}=100$ and $300$.

We expect the off-axis counterpart of the prompt gamma-ray emission to be longer-lived and fainter. We find that, for a slightly off-axis viewing angle $\theta_{\rm v}=6^{\circ}$ (near the jet edge), the non-thermal prompt emission remains brighter than $\sim\!10^{46}{\rm \,erg\,s^{-1}}$ at $\sim$300\,s, which exceeds the thermal emission from cocoon cooling, and thereby dominates the observation while simultaneously hardening the spectrum.

\subsection{Underprediction of the UV/Optical Luminosity}
\label{sec:lumino}
We discuss several potential solutions for the underprediction of the UV/optical luminosity below. Testing these ideas is beyond the scope of this work.

(1) Increasing the energy budget. According to Eq. (\ref{eq:Lbol}), the bolometric luminosity from cocoon cooling scales linearly to the initial thermal energy of the cocoon and also the radius of the progenitor star, $L_{\rm bol}\propto E_{\rm c,th0}R_{\star}$. A more energetic cocoon will therefore produce brighter emission than our results. \cite{Zhu2025MNRAS.544L.139Z} fit the cooling emission of EP250108a, and infer an outer cocoon energy of $\sim10^{52}{\rm erg}$ and a stellar radius of $R_{\star}\simeq 5R_{\odot}$. However, such a large outer cocoon energy may not be typical for LGRBs, as the beaming-corrected jet energy is often of the order $10^{51}{\rm \,erg}$ \citep{Cenko2010ApJ...711..641C,Shibata2025PhRvD.111l3017S}.

(2) Increasing the (effective) stellar radius. Here, we consider the effective stellar radius to be where the outer cocoon begins free expansion.
If there is some circumstellar medium (CSM) around the star, the effective stellar surface may be greatly enhanced, potentially reaching $R_{\rm CSM}\sim100R_{\star}$ \citep{Nakar2015ApJ...807..172N}. The CSM model has been proposed to explain the FXT and low luminosity GRBs \citep{Hamidani2025ApJ...986L...4H}. So far, there has been no independent confirmation of such extended CSM around LGRB progenitors.

(3) Introducing additional outflow components. We only include the jet power in this work and there could be other powerful outflows within the collapsar. One candidate is the disk wind launched by the central engine \citep[e.g.][]{Gottlieb2025ApJ...992L...3G}. Previous works suggested that the disk wind carries $\sim10^{52}{\rm erg}$ energy and likely some heavy elements that can power thermal emission \citep[e.g.,][]{2005ApJ...629..341K, 2023ApJ...956..100F, Shibata2025PhRvD.111l3017S}. 

As for the FXT signal, the X-ray luminosity will be higher in the case with increased energy and a larger stellar radius, and the introduction of additional non-relativistic disk wind does not affect the X-ray emission from the mildly relativistic inner cocoon.

\section{Summary}
\label{sec:sum}
In this paper, we perform numerical simulations of the jet-cocoon system in LGRBs from the jet launching to the cocoon's diffusion radius ($\sim\!10^{14}\rm\, cm$). We also develop a new post-processing framework to compute the observed cocoon cooling emission at different viewing angles, taking into account photon diffusion and relativistic effects.

Our main result is that, for viewing angles $\theta_{\rm v}=10^{\circ}$-$20^{\circ}$, the off-axis cocoon emission can produce FXTs with luminosity $L_{\rm X}\simeq 10^{47-48} {\rm\, erg\,s^{-1}}$ and duration $t_{\rm X}\simeq 10$-$100\,$s. The observed spectra are quasi-thermal with the peak energy $E_{\rm peak}\simeq 0.8$ keV. These properties naturally explain observational features of a fraction of FXTs detected by EP WXT, including the high luminosity, soft spectra, and the lack of gamma-ray counterparts. Both the observed luminosity and duration of the X-ray emission decrease with increasing viewing angle. At $\theta_{\rm v}=45^{\circ}$, the peak X-ray luminosity is $\sim\!10^{46}{\rm erg\,s^{-1}}$ and the duration is only 10\,s, making WXT detection feasible only for nearby events ($z\lesssim 0.05$).


The X-ray spectra of cocoon cooling emission are generally much softer than the non-thermal emission from GRB jets and they become softer over time. The photon indices of $T_{90}$-averaged spectra in all models are softer (i.e., larger) than 2.5. 
This spectral softness is caused by the relatively low color temperature of the emission: $T_{90}$-averaged peak energy $E_{\rm peak}\simeq0.8$ keV for all models and hence most photons are emitted below EP WXT band (0.5-4 keV). The spectral softening during the FXT phase is caused by the rapid expansion and cooling of the cocoon, with the peak energy decaying roughly as a power-law $E_{\rm peak}\propto t_{\rm obs}^{-0.6}$. These properties provide useful diagnostics for FXTs originating from LGRBs.

The Rayleigh-Jeans tail of the X-ray emission also produces a simultaneous UV flash with peak luminosity of the order $\sim\!10^{43}{\rm\, erg\,s^{-1}}$, for $\theta_{\rm v}\leq20^{\circ}$. At later times (hours to days), the cooling emission shifts to the UV and optical bands and produces an optical plateau lasting for a few days. We note that our current model underpredicts the luminosity of the optical plateau associated with FXTs and SNe Ic-BL on the timescale of 1 day. This discrepancy is likely due to missing additional (sub-relativistic) outflow components and/or larger effective stellar radii than in our models.

The jet and cocoon both produce non-thermal afterglow emission, which may dominate the X-ray emission following the FXT phase and contribute to the optical emission (see \cite{Zheng2025ApJ...985...21Z} for detailed discussions of afterglow).

\begin{acknowledgments}
We thank Jin-Ping Zhu, Bing Zhang, Ore Gottlieb, Ryan Foley, and Xiang-Yu Wang for helpful discussions.  J.-H.Z.’s work is supported by the National Natural Science Foundation of China (grant numbers 124B2057, 13001075).
\end{acknowledgments}

\appendix
\renewcommand{\thefigure}{A\arabic{figure}}
\setcounter{figure}{0}
\section{Convergence Test}
\label{app:conver}

\begin{figure}
    \centering
    \includegraphics[width=0.45\linewidth]{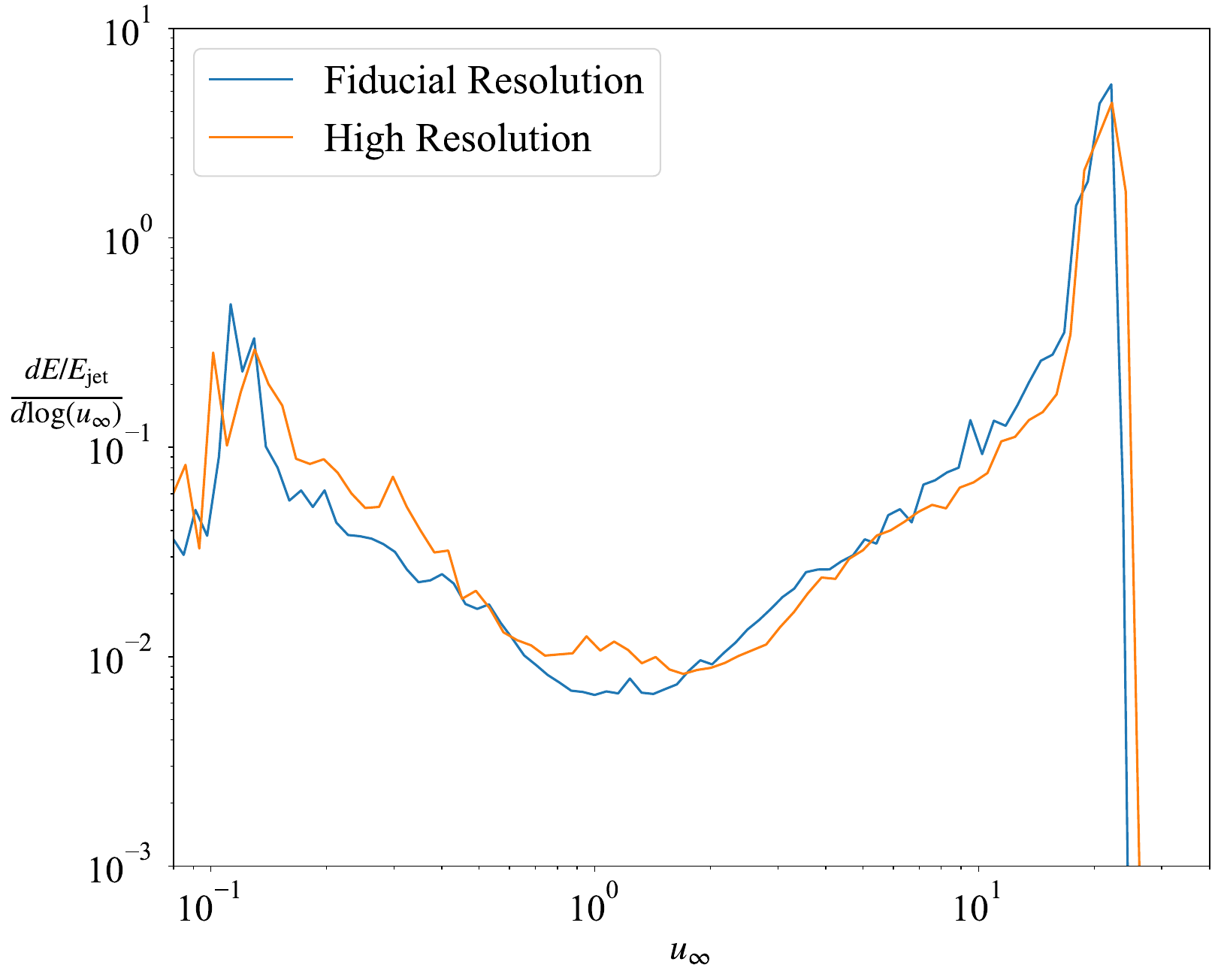}
    \includegraphics[width=0.45\linewidth]{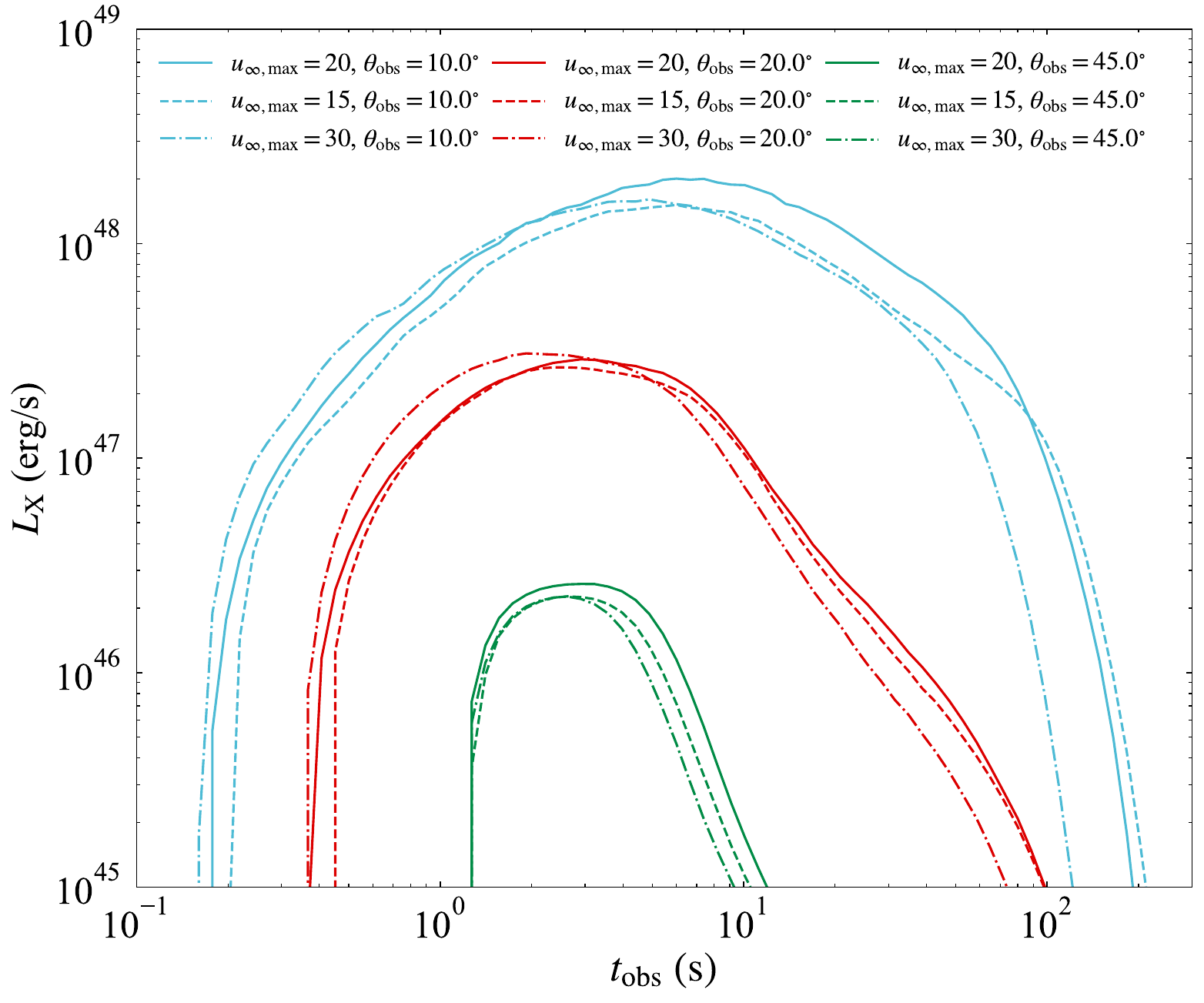}
    \caption{\textit{Left panel}: The energy distribution per logarithm of the terminal 4-velocity. The blue line shows the $Lc$ model at the fiducial resolution, while the orange line shows the same model at higher resolution. Both distributions are computed at $t_{\rm lab}=40$ s when the jet head is reaching $\sim10R_{\star}$. 
    \textit{Right panel}: EP WXT X-ray lightcurves. The solid lines are the fiducial lightcurves with $u_{\rm\infty,max}=20$ in $Lc$ model that we show in the left panel of Figure \ref{fig:Lc}. The dashed lines and dash-dotted lines are lightcurves of $Lc$ model with $u_{\rm\infty,max}=15$ and $u_{\rm\infty,max}=30$, respectively. Colors indicate different viewing angles, as in the left panel of Figure \ref{fig:Lc}.
    }
    \label{fig:reso}
\end{figure}

We verify that our results are insensitive to our simulation setups.

The first test is for the spatial resolution. We carried out higher-resolution simulations of the \textit{Lc} model with doubled grid points on both radial and angular dimensions. The difference between the jet breakout times is $\sim\!1$ second, indicating the cocoon energies $E_{\rm c}\approx L_{\rm j}(t_{\rm bo}-R_{\star}/c)$ are similar. Furthermore, we compare the energy distributions over the terminal 4-velocity in the left panel of Figure \ref{fig:reso}. The higher-resolution energy distribution is consistent with results obtained in the fiducial resolution used in the main text. Since the luminosity is primarily determined by the total cocoon energy, and the lightcurve shape is determined by the energy-velocity distribution, we conclude that our lightcurves have reached convergence.


The second test examines the impact of the maximum terminal 4-velocity $u_{\rm\infty,max}$.
We calculate the lightcurves based on two additional simulations of the \textit{Lc} model using $u_{\rm\infty,max}=15$ and $30$, as opposed to the fiducial choice of 20. The lightcurves for different viewing angles $10^\circ \leq \theta_{\rm v}\leq 45^\circ$ are shown in the right panel of Figure \ref{fig:reso}. We find the lightcurves to be broadly consistent with the canonical model we present in the main text. This is because the fluid elements near the line of sight ($\theta_{\rm v} \geq 10^{\circ}$) are only mildly relativistic ($u_{\infty}\lesssim 10$), so the maximum Lorentz factor within the jet core does not significantly affect the off-axis emission.

\section{Local thermodynamic equilibrium before the Breakout}
\label{app:LTE}
In the main text, we assume that the radiation field in the jet-cocoon system is described by a blackbody in the comoving frame. This assumption is not always valid in relativistic jets because the gas is dilute and potentially unable to produce enough photons to thermalize the radiation within the dynamical expansion timescale. We define the ratio between the internal energy density and the free-free emission over the dynamical time in the comoving frame, which is given by
\begin{equation}
    \label{eq:LTE}
    \eta\equiv\frac{u'_{\rm int}}{\epsilon'_{\rm ff}t'_{\rm dyn}},
\end{equation}
where $u'_{\rm int}=3p'$ is the internal energy density, $\epsilon'_{\rm ff}=1.4\times10^{-27}\sqrt{T'}n'^2 {\rm erg\,cm^{-3}\,s^{-1}}$ is the free-free energy production rate, $n'$ is the gas number density, $T'=(u'_{\rm int}/a)^{1/4}$ is the blackbody equivalent temperature, and $t'_{\rm dyn}=r/(\Gamma\beta c)$ is the dynamical timescale. The parameter $\eta$ greater than 1 would mean that the thermalization by free-free emission/absorption is insufficient and the system is in the ``photon-starved'' regime that deviates from local thermodynamic equilibrium (LTE). On the other hand, if $\eta < 1$, we know that the radiation field near a given fluid element would reach LTE and, during the subsequent expansion of the cocoon, the radiation spectrum would remain the blackbody shape.

\begin{figure}
    \centering
    \includegraphics[width=0.5\linewidth]{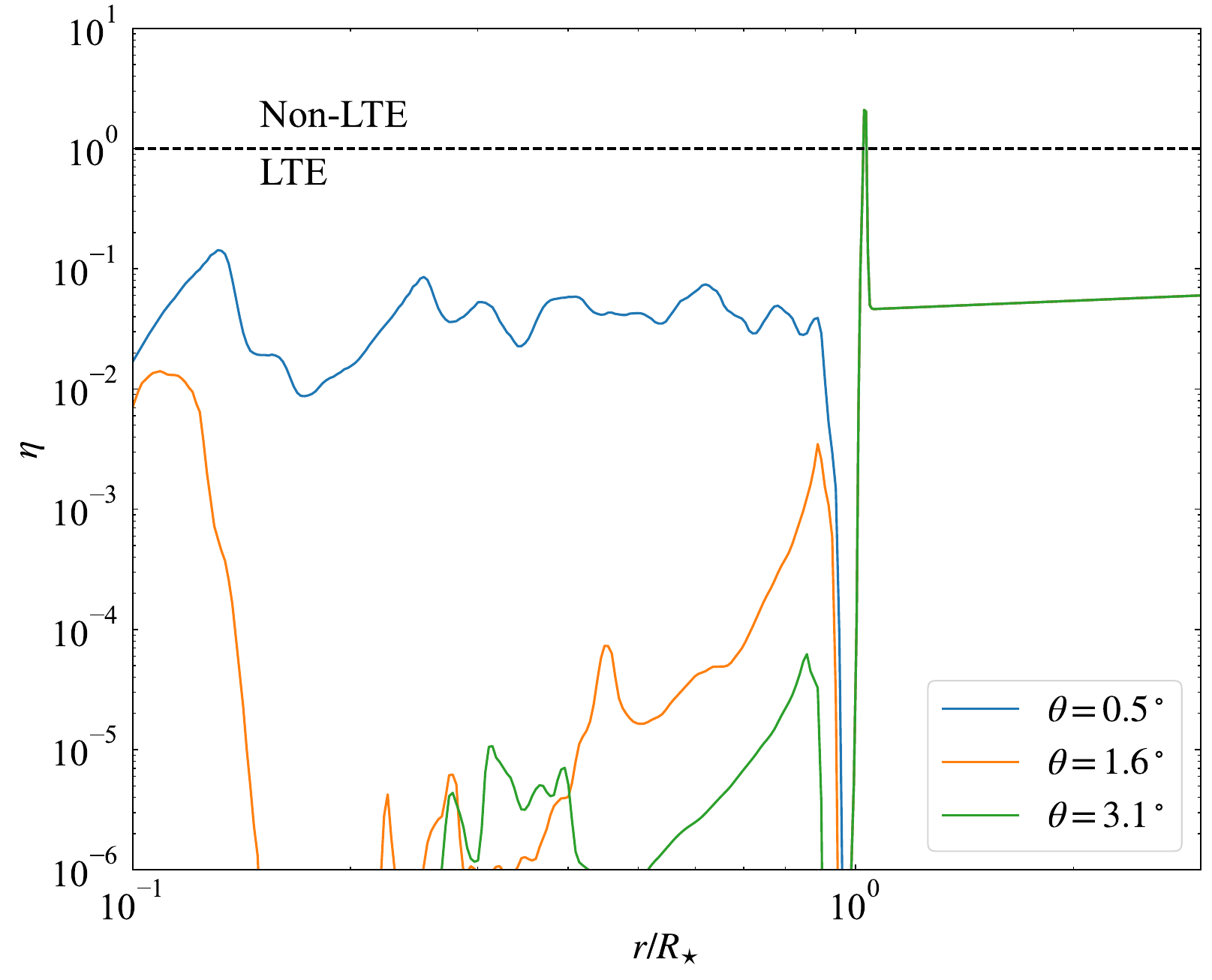}
    \caption{The ratio $\eta$ (Eq. \ref{eq:LTE}) between energy density and free-free energy production. The blue ($\theta=0.5^{\circ}$), orange ($\theta=1.6^{\circ}$), and green ($\theta=3.1^{\circ}$) lines are for the fluid elements along different polar angles. The black dashed line is the threshold for the LTE condition $\eta\leq 1$. Almost all fluid elements of the jet-cocoon system are in the LTE regime except for the layer near the stellar surface (the gas near and outside the stellar surface is unimportant for the subsequent cocoon evolution).
    }
    \label{fig:LTE}
\end{figure}

The energy density $u_{\rm int}'$, gas number density $n'$, and velocity $\beta$ can be obtained from the simulation.
Figure \ref{fig:LTE} shows $\eta$ for different fluid elements within the cocoon at the breakout time. We find that nearly all fluid elements are well in the LTE regime with $\eta\lesssim 0.1$. This includes even the gas on the jet axis (see the $\theta=0.5^\circ$ case). The reason is that we set the maximum terminal Lorentz factor to be 20 in the simulation. For a more realistic LGRB jet with $\Gamma_{\rm j}\gtrsim 300$, this region is likely to be in non-LTE because $\eta\propto\Gamma^2$ is very sensitive to the Lorentz factor \citep{Zheng2025ApJ...985...21Z}. The fluid elements in the shock-heated cocoon are orders of magnitude away from the threshold of $\eta = 1$ due to the lower Lorentz factors. Therefore, we conclude that the majority of the jet-cocoon fluid elements are in the LTE regime before the breakout. 

{
We emphasize that Eq. \ref{eq:LTE} is used to judge whether LTE can be established after the shock heating, rather than as a condition that must be satisfied at every time. After the breakout, the cocoon expands adiabatically without additional strong heating. Therefore, if LTE is reached before breakout, the radiation spectrum will remain close to a blackbody subsequently.
}

\section{The integral range of off-axis cocoon}
\label{app:sphere}
{
In this appendix, we show that, for our 2D axisymmetric simulations, integrating the jet-cocoon emission over the northern hemisphere (the one closer to the observer) as done in Eq. (\ref{eq:observed_flux}) is a good approximation to obtain the observed flux.
}

{
We first consider the non-relativistic case, where the geometry is simpler. For an optically thick sphere (e.g., the Sun), a distant observer receives radiation only from the visible hemisphere, namely the set of surface elements whose outward normal has a positive projection along the line of sight (i.e., $\cos\chi>0$).  The observed specific flux is therefore $F_{\nu}=\int_{\cos\chi>0} I_{\nu}\cos\chi d\Omega$. The integral range is the visible surface whose projection factor is positive (i.e. $\cos\chi>0$). 
}

{
If the radiation is isotropic in the lab frame, we may choose the observer's line of sight as the polar axis. In that case, the projection factor $\cos\chi$ reduces to $\cos\theta$ (i.e. $\theta_{\rm v}=0$ in Eq. \ref{eq:project}) and the integral range is simply the northern hemisphere (i.e. $0<\theta<\pi/2$) because the observer is on the north pole. The above expression then reduces to the familiar result $F_{\nu}= \int^{\pi/2}_0 I_{\nu}\cos\theta d\Omega=\pi I_{\nu}$. }

{
For the cocoon, however, the intensity depends on polar angle $I_{\nu}(\theta)$.
In principle, one should therefore integrate over the actual visible surface, which changes with the viewing angle. For our 2D simulations, however, this can be simplified because the system has both axial symmetry and equatorial symmetry. The key point is that the part of the southern hemisphere that is visible to the observer is paired with the part of the northern hemisphere that is invisible. We denote these two regions as Regions 3 (visible surface in the southern hemisphere) and 1 (invisible part of the northern hemisphere), respectively, while Region 2 is the overlapping part shared by the visible surface and the northern hemisphere.
Because of the two symmetries, every cell in region 1 has a corresponding “twin” cell in region 3, related by $\theta_3=\pi-\theta_1$ and $\phi_3=\pi+\phi_1$. These two cells have the same physical properties and the same solid angle, but their projection factors satisfy $\cos\chi_3=-\cos\chi_1$ (see Eq.\ref{eq:project}). 
}

{ 
Because of the two symmetries, every cell in Region 1 has a corresponding ``twin'' cell in Region 3, related by $\theta_3=\pi-\theta_1$ and $\phi_3=\pi+\phi_1$. These two cells have the same physical properties and the same solid angle, but their projection factors satisfy $\cos\chi_3=-\cos\chi_1$ (see Eq.~\ref{eq:project}).
}

{
As a result, in the non-relativistic limit, integrating over the true visible surface is exactly equivalent to integrating over the northern hemisphere with $|\cos\chi|$:
\begin{equation}
    \int_{\cos\chi>0} I_{\nu}(\theta)\cos\chi d\Omega=\int^{\pi/2}_0\int^{2\pi}_0 I_{\nu}(\theta)|\cos\chi| d\Omega.
\end{equation}
In the relativistic case, the only difference comes from the Doppler boosting $I_{\nu}=\mathcal{D}^3I'_{\nu'}$ because the Doppler factor also depends on the projection factor $\mathcal{D}=(\Gamma(1-\beta\cos\chi))^{-1}$. Two twin cells with opposite signs of $\cos\chi$ no longer contribute equally.
The visible southern cell has $\cos\chi>0$, and therefore has a larger Doppler factor than its invisible northern counterpart, which breaks the equality above. However, in practice, this correction is very small in our simulations. We directly compare the results from two integrations and find that the maximum difference in our lightcurves is less than $2\%$. 
This is because any ultra-relativistic material in Region 3 contributes negligible observed flux as the emission is strongly beamed away from the LOS.}



\bibliography{reference}{}
\bibliographystyle{aasjournal}



\end{document}